\documentclass[useAMS,usenatbib]{mn2e}

\usepackage{epsfig,graphicx,graphics}

\title[Cm-wavelength continuum in M~78]{Dust-correlated
  centimetre-wave radiation from the M~78 reflection nebula}
\author[Castellanos et al.]{Pablo Castellanos$^1$, Simon
  Casassus$^{1,2}$\thanks{E-mail:
simon@das.uchile.cl}, Clive Dickinson$^3$,
  Mat\'{\i}as Vidal$^1$,\newauthor Roberta Paladini$^4$,
  Kieran Cleary$^4$, Rodney D. Davies$^3$,\newauthor
  Richard J. Davis$^3$, Glenn J. White$^{5,6}$, Angela
  Taylor$^7$\\
$^{1}${Departamento de Astronom\'{\i}a, Universidad de
  Chile, Casilla 36-D, Santiago, Chile}\\
$^{2}${Observatoire de Paris, LUTH and Universit\'e Denis Diderot, Place J. Janssen, 92190 Meudon, France}\\
$^{3}${Jodrell Bank Centre for Astrophysics, Alan Turing
  Building, School of Physics \& Astronomy,}\\ {The University of
  Manchester, Oxford Road, Manchester, M13 9PL, UK}\\
$^{4}${
Spitzer Science Center, California Institute of Technology, Pasadena,
CA. 91125.}\\
$^{5}${The Rutherford Appleton Laboratory, Didcot,
  Oxfordshire OX11 0QX, UK}\\
$^{6}${Department of Physics \& Astronomy, The Open University, Milton
      Keynes, MK7 6AA, UK}\\
$^{7}${Denys Wilkinson building, Physics Department, Oxford
University, Keble Road, Oxford OX1 3RH}\\
}

\begin{document}

\date{Accepted 1988 December 15. Received 1988 December 14; in original form 1988 October 11}

\pagerange{\pageref{firstpage}--\pageref{lastpage}} \pubyear{2002}

\maketitle

\label{firstpage}

\begin{abstract}
An anomalous radio continuum component at cm-wavelengths has been
observed in various sources, including dark clouds. This continuum
component represents a new property of the ISM.  In this work we focus
on one particular dark cloud, the bright reflection nebula M~78. The
main goal of this work is to investigate the cm-wave continuum
emission in a prominent molecular cloud, nearby and with complementary
observational data. We acquired Cosmic Background Imager (CBI)
visibility data of M~78 at 31~GHz with an angular resolution of $\sim
5.8\arcmin$, and CBI2 data at an angular resolution of $\sim
4.2\arcmin$. A morphological analysis was undertaken to search for
possible correlations with templates that trace different emission
mechanisms. Using data from {\em WMAP} and the Rhodes/HartRAO 2326~MHz
survey we constructed the spectral energy distribution (SED) of M~78
in a 45$\arcmin$ circular aperture. We used results from the
literature to constrain the physical conditions and the stellar
content. The 5~GHz -- 31~GHz spectral index in flux density ($\alpha =
1.89\pm0.15$) is significantly different from optically thin-free-free
values.  We also find closer morphological agreement with IR dust
tracers than with free-free sources.  Dust-correlated cm-wave emission
that is not due to free-free is significant at small scales (CBI
resolutions). However, a free-free background dominates at
cm-wavelengths on large scales ($\sim 1~$deg).  We correct for this
uniform background by differencing against a set of reference fields.
The differenced SED of M~78 shows excess emission at 10-70~GHz over
free-free and a modified blackbody, at 3.4~$\sigma$. The excess is
matched by the spinning dust model from \citet{dra98}.

\end{abstract}


\section{Introduction}

Since 1996, experiments designed to measure the CMB anisotropy have
reported an anomalous diffuse foreground in the range of
10-60~GHz. This diffuse emission is correlated with thermal emission
from dust grains at 100~$\mu$m. The spectral index (considering $S_\nu
\propto \nu^\alpha$) of the radio-IR correlated signal is
$\alpha_\mathrm{radio/IR} \sim 0$ in the range 15-30~GHz, as for
optically thin free-free \citep{kog96}. But $\alpha_\mathrm{radio/IR}
\sim -0.85$ between 20-40~GHz, for high-latitude cirrus \citep{dav06}.
Additionally the absence of H$\alpha$ emission concomitant to radio
free-free emission, would require an electron temperature $T_e
\geqslant 10^6$ to quench H\,{\sc i} recombination lines
\citep{lei97}. Another emission mechanism was presented by
\citet{dra98}, who calculated that spinning interstellar dust grains
produce rotational dipole emission in the range from 10 to 100 GHz, at
levels comparable to the excess diffuse foreground observed over the
free-free component.

Observations of specific targets may shed light on the anomalous
foreground, whose existence in the diffuse ISM has been  inferred
statistically. Anomalous cm-wavelength radiation has been found in
about a dozen clouds \citep[][]{fin04, wat05, cas06, cas08, sca09a,
  dic09, vid10, sca09b}.

The search by \citet{fin02} for anomalous cm-wave emission in 9 dark
clouds, resulted in one detection: LDN~1622\footnote{and also another
  detection, which was later refuted by \citet{dic06}}. The radio
spectrum of LDN~1622 matches that of spinning dust emission
\citep[][]{fin04}.  Cosmic Background Imager \citep[CBI,][]{pad02}
observations of LDN~1622 linked, on morphological grounds, the cm-wave
emitters to the smallest interstellar dust grains (i.e. the VSGs) - as
required for spinning dust \citep[][]{cas06}. Interestingly, LDN~1622
is notable in being exposed to UV radiation from the adjacent hot
stars in the Ori~OB~1b association.

The prototypical dark cloud LDN~1688, close to the $\rho$~Oph star,
was recently found to be bright at cm-wavelengths \citep[][]{cas08}.
The $\rho$~Oph cloud is undergoing intermediate-mass star
formation. UV radiation from its hottest young stars heats and
dissociates exposed layers, but does not ionize hydrogen. Only faint
radiation from the Rayleigh-Jeans tail of $\sim$~10--100~K dust is
expected at wavelengths longwards of $\sim$3~mm. Spinning dust
comfortably explains the radio spectrum of $\rho$~Oph~W, as in
LDN~1622. However, spinning dust encounters difficulties in explaining
the cm-wave morphology of the $\rho$~Oph cloud. Diffuse mid-IR
(10--20~$\mu$m) emission from $\rho$~Oph is interpreted as stochastic
heating of VSGs \citep[][]{ber93}. Dust emissivities are proportional
to both the dust grain density and the local UV energy
density. Spinning dust was introduced to explain the 100~$\mu$m--1~cm
correlation in the diffuse ISM, but does this correlation extend to
smaller scales in denser environments?  \citet{cas08} report that the
mid-IR intensity peaks in $\rho$~Oph, from the circumstellar nebulae
about Oph~S~1 and Oph~SR~3, have no 31~GHz counterparts. They find
that the predicted spinning dust intensities towards S~1 are in excess
over the observed values by a factor of $>40$.  On the basis of {\em
  Spitzer}~IRS infrared spectroscopy, this discrepancy could
marginally be explained by VSG depletion near S~1 and SR~3.

As part of an effort to understand the emission mechanisms in
environments giving rise to the 31 GHz continuum, we have studied a
dark cloud with extensive complementary data: the reflection nebula
M~78, with good constraints on its physical environment. Several
observations are available in the literature that constrain the
physical conditions in M~78, making it a promising testbed of
candidate emission mechanisms. As a reflection nebula M~78 has
relatively low levels of free-free emission but also contains stars
bright enough to excite circumstellar dust grains. Reflection nebulae
are regions of star formation, and they contain large amounts of dust.

\begin{figure}
\begin{center}
\includegraphics[width=\columnwidth,height=!]{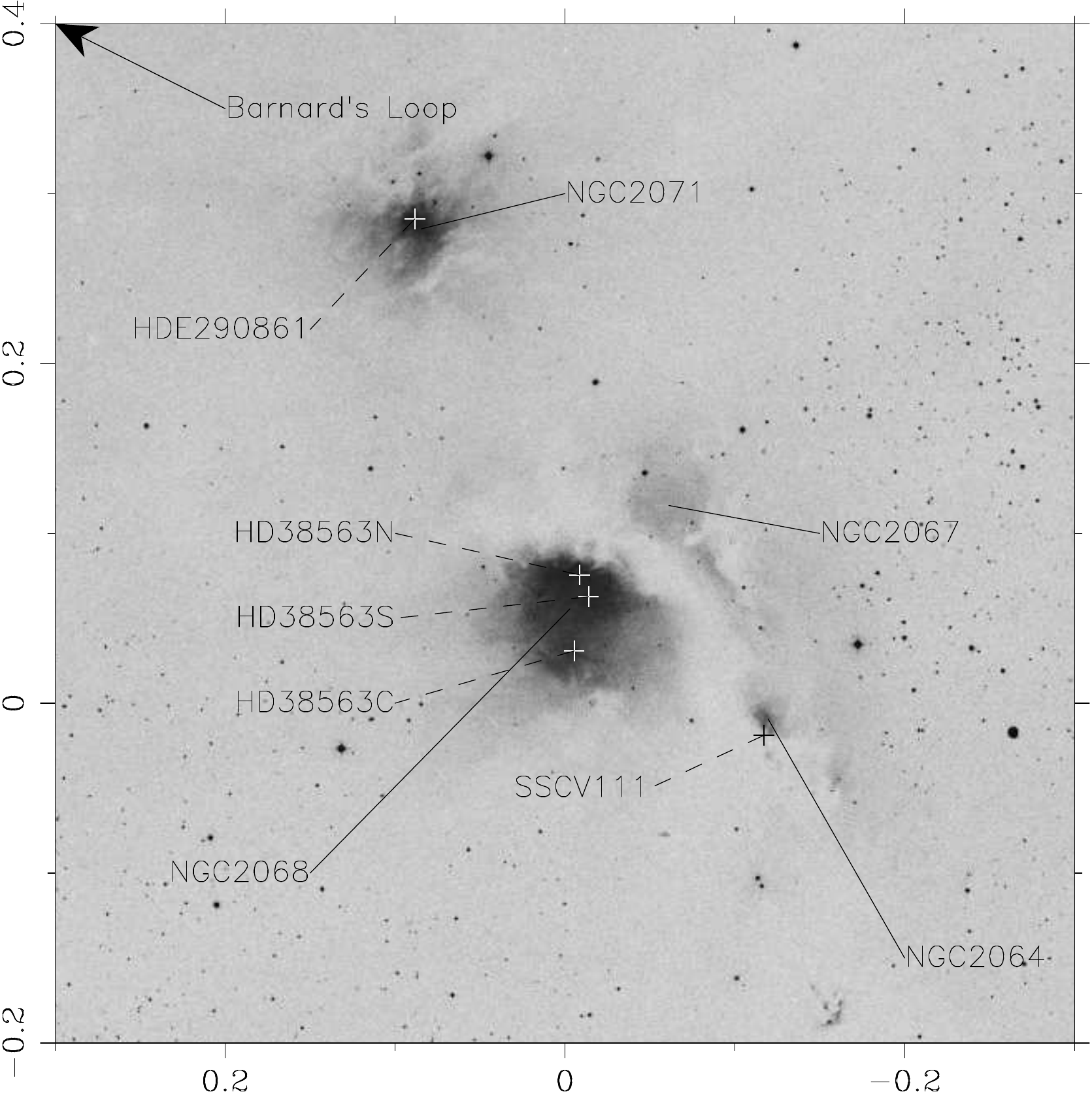}
\end{center}
\caption{\label{figure:posi}M~78 area taken from {\it DSS} blue band
  Survey. The largest and brightest nebula is NGC~2068/M~78, the
  arc-shaped nebula to the west is NGC~2067. Northeast from NGC~2068
  is NGC~2071 and the faintest nebula, southwest from NGC~2068 is
  NGC~2064. The crosses indicate the positions of the brightest stars
  in the field. The dashed lines point to these stars and the nebulae are
  pointed to with solid lines. The coordinates are in right ascension and
  declination offsets from ($05^h46^m46^s.7 +00\degr 00\arcmin 50\arcsec$
  J2000), in degrees of arc.}
\end{figure}

In this report we refer to M~78 as a group of reflection nebulae, all
of them part of LDN~1630. This is a large dark cloud in the
constellation of Orion, north from the belt, which also contains
NGC~2023.  LDN~1630 corresponds to the Orion B molecular cloud, which
has recently been studied in molecular lines by \citet[][]{buc10}. The
area that hereafter will be called M~78 contains the following
nebulae: {NGC 2068} (sometimes also referred to as {M 78}), {NGC
  2071}, {NGC 2064} and {NGC 2067} (see Figure~\ref{figure:posi}).

Section~\ref{sec:m78} refers to the physical conditions and different
phases of gas and dust found in M~78. In Sec.~\ref{sec:obs} we will
describe the observations used for this work. Sec.~\ref{sec:obs} also
includes a discussion about the CBI flux loss.  In Sec.~\ref{sec:disc}
we present our results, derived from a comparison between the CBI data
and various tracers of dust and free-free emission. In
Sec.~\ref{sec:conc} we summarize our conclusions, quantifying the
31~GHz radiation from M~78.


\section{Environment and Physical Conditions}\label{sec:m78}

\begin{figure*}
\begin{center}
\includegraphics[width=\textwidth,height=!]{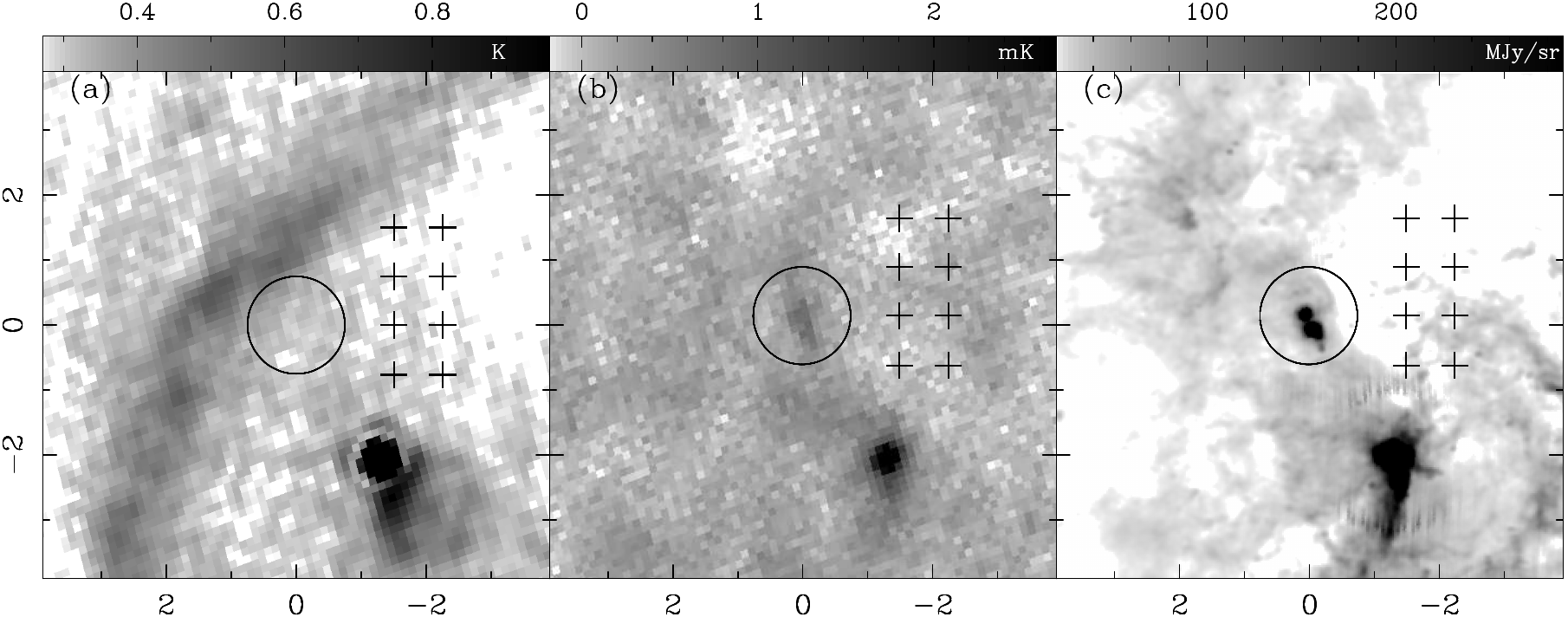}
\end{center}
\caption{\label{figure:large}Large scale structure around M~78 in
  various frequencies. The solid line circle is the aperture where we
  integrated to extract flux densities. The crosses indicate the
  center of the reference fields used for differentiation. The bright
  source at $\sim 2.5$ degrees southwest from M~78, common to each
  image, is NGC~2023. (a) At 2326~MHz, from Rhodes/HartRAO survey, we
  see clearly the arc to the northeast which corresponds to part of
  the Barnard's Loop, but there is no structure at the position of
  M~78. (b) At 93.4~GHz, corresponding to {\em WMAP} W band, we see a
  faint feature corresponding to M~78.(c) Image at 100~$\mu$m from
  IRIS, we see that the structure of M~78 corresponds to that observed
  in (b). $x-$ and $y-$axis show right-ascension and declination
  offsets from ($05^h46^m50^s.4 +00\degr 09\arcmin 31\arcsec$ J2000),
  in degrees of arc in degrees of arc.}
\end{figure*}

L1630 is embedded in the Orion Complex, at a distance of about 400~pc
\citep{ant82}. It is part of Orion Molecular Cloud II, among other
star forming regions and recently formed clusters. We also find
Barnard's Loop in this region; it passes at $\sim 40\arcmin$ from the
northern end of NGC~2071. As a reflection nebula M~78 has low levels
of free-free emission but is exposed to stars bright enough to excite
IR emission from dust. Reflection nebulae are regions of star
formation, so they contain young stars and dense molecular cores.  As
we can see in Fig.~\ref{figure:large}, although the large scale radio
emission is dominated by a uniform free-free background, the
individual features of M78 become more pronounced with increasing
frequency and angular resolution.


The average H-nucleus density value in L1630, derived from CO low
resolution maps, is $n \sim 10^3$~cm$^{-3}$, but towards NGC~2068 and
NGC~2071, higher densities are found \citep{str75}.  \citet{lad97}
obtained, for these clumps, densities of $\log(n/\mathrm{cm^{-3}}) =
5.3$ for NGC 2068 and $\log(n/\mathrm{cm^{-3}}) = 5.9$ for NGC
2071. These values were derived from an excitation analysis of CS line
fluxes for three transitions $J = 2 \rightarrow 1$, $J = 3 \rightarrow
2$ and $J = 5 \rightarrow 4$. \citet{buc10} have recently reported on
a large scale CO(3-2) isotopologue survey of Orion B, including the
M~78 area, and focusing on the star formation activity. Depending on
the isotopologue, the molecular mass of NGC~2071 is 0.4 to
3.7~10$^{3}$~M$_\odot$. NGC~2071 dominates NGC~2068 by mass.

M~78 is still undergoing star formation \citep{fla08}. M~78 contains
two embedded clusters, in both NGC~2071 and NGC~2068. These clusters
are considered to have an age of about $2 \pm 1$~Myr, and for a
significant fraction of stars in the cluster there is evidence of
accretion disks.

The illuminating stars for {NGC 2068} and NGC~2071 have been
described by \citet{str75}. NGC~2068 hosts the star HD38563N (B2\,{\sc
  iii}) and embedded in NGC~2071 is HDE290861 (B2-B3). Other early
type stars in NGC~2068 are HD28563S (B3-B5) and HD38563C (A0\,{\sc
  ii}). These stars appear to be very young, with ages around $10^5$
years. For {NGC 2064} the illuminating star is SSCV~111
\citep{str75}, and its spectral type is B3\,{\sc v} \citep{chi84}. 

The photo-dissociation-region (PDR) surrounding HD38563N was studied
by \citet{owl02}, finding a dust temperature of 41~K, a gas density of
5000~cm$^{-3}$ and a gas temperature of 250~K. They also give $G_\circ
= 2800$ for the UV radiation incident on NGC~2068, where $G_\circ = 1$
corresponds to the average interstellar radiation field of
$1.6~10^{-3}$~erg cm$^{-2}$ s$^{-1}$ \citep{mat83}.

M~78 has an associated C\,{\sc ii} region. According to \citet{pan78}
the C\,{\sc ii} region extends $\sim 20\arcmin$ along a line between
NGC~2071 and NGC~2068. The electron temperature ranges from 20~K to
50~K and the electron density is 0.2-1.0 cm$^{-3}$
\citep{bro75,pan78}. Both densities and temperatures were derived from
the ratio of the power in two carbon recombination lines. According to
\citet{sil83} this C\,{\sc ii} region is unusual because it does not
present a sulfur line in its spectrum. This sulfur line is found
towards all other known C\,{\sc ii} regions. Another datum is that the
carbon radio recombination line is the narrowest yet observed for
reflection nebulae and has fainter brightness temperatures compared to
other C\,{\sc ii} regions.

Besides the C\,{\sc ii} regions, a compact H\,{\sc ii} region
($\sim 1\arcmin$) was reported by \citet{mat76} around the
brightest star in NGC 2068, HD38563N. \citet{mat76} measured 0.1~Jy at
2.4~GHz. By assuming a temperature of $\sim$ 7000 K, they found an
electron density of 40 cm$^{-3}$. This region is responsible for most
of the free-free emission of the nebula on 1-10$\arcmin$ scales,
although on 1~deg scale it is weaker than the background diffuse
H\,{\sc ii} region.

M~78 also hosts H$_2$O masers \citep{cam78}. These masers are
considered tracers of proto-stellar disks, given that they coincide
with infrared sources with associated IR excesses (i.e. circumstellar
disks) and collimated outflows (e.g. as in class-II protostellar
objects).

\section{Observations} \label{sec:obs}

\subsection{Cosmic Background Imager data}


The CBI is described in \citet{pad02}. The CBI is a planar
interferometer array with 13 antennas, each 0.9~m in diameter, mounted
on a 6~m tracking platform. The CBI receivers operate in 10 frequency
channels covering 26--36~GHz.  It is located in Llano de Chajnantor,
Atacama, Chile. Cancellation of ground and Moon contamination was obtained
by differencing with a reference field at the same declination but
offset in hour angle by the duration of the on-source integration. We
used an on-source integration time of 8~min, with a trailing reference
field.

\begin{figure*}
\begin{center}
\includegraphics[width=0.8\textwidth,height=!]{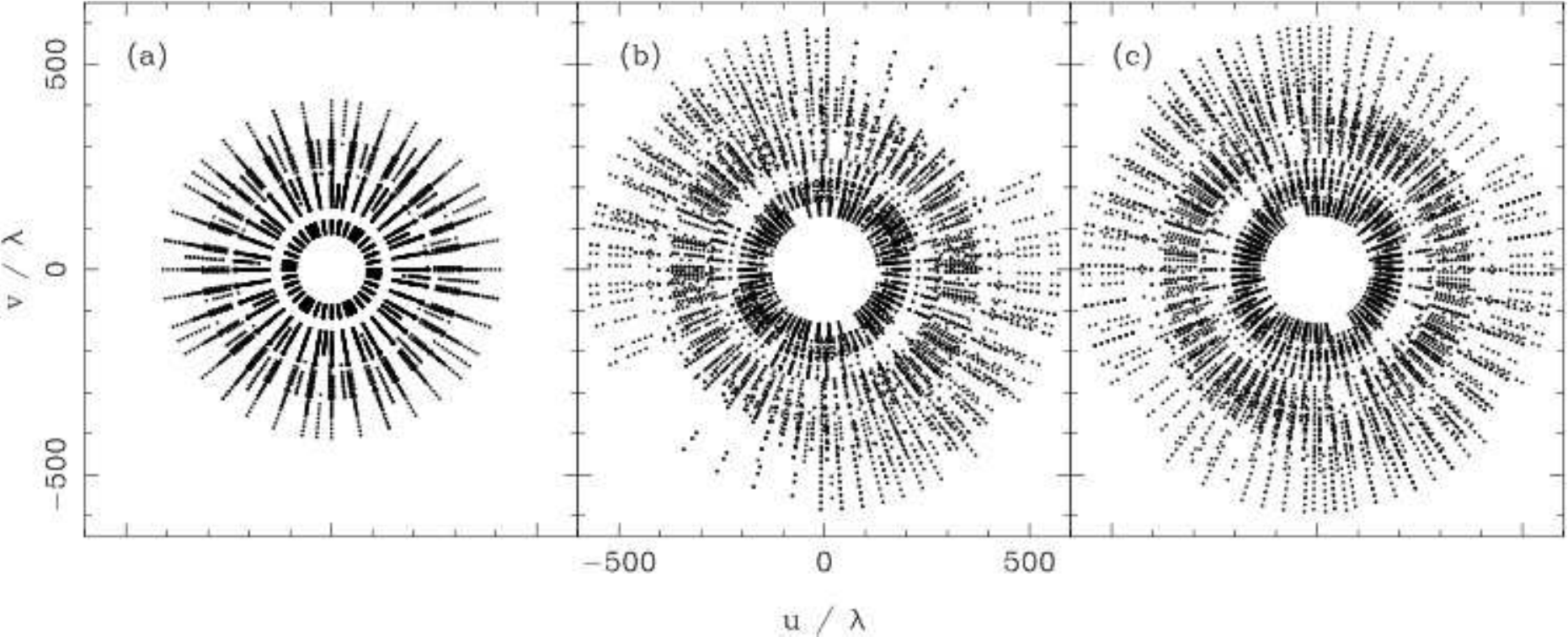}  
\end{center}
\caption{\label{figure:uv}The $uv$ coverage for (a) CBI when taking M~78, (b)
  for the observation of NGC~2068 with CBI2 and (c) for CBI2 in
  NGC~2071.}
\end{figure*}

We used the CBI to observe M~78 on 20 December 2004. The primary beam
is 45$\arcmin$.2 at FWHM of the best fitting Gaussian at the reference
frequency of 31~GHz, and the synthesized beam is $5\arcmin.94 \times
5\arcmin.70$ with uniform weights. We also observed M~78 with CBI2, an
upgrade where the 0.9~m CBI dishes were changed for 1.4 m dishes. In
the case of CBI2, since the primary beam FWHM is 28$\arcmin$.2 (this
beam diverges from a Gaussian outside $\sim$FWHM$/2$), we mosaiced two
pointings on NGC 2071 and NGC 2068. The $uv$ coverage was different in
the two fields, thus they have different synthesized beams (using
uniform weights); for NGC 2068 we have $4\arcmin$$.89 \times
3\arcmin$$.97$, while for NGC 2071 the values are $4\arcmin$$.37
\times 4\arcmin$$.00$. The visibility coverage in the $uv$ plane is
displayed in Fig.~\ref{figure:uv}.

The data were reduced and edited using a special-purpose package
(CBICAL, developed by T.J. Pearson).  Flux calibration was performed
using Tau~A, whose fluxes are in turn calibrated against Jupiter
\citep[with a temperature of $146\pm0.75$~K,][]{hil09}. The flux
calibrator is also used as the reference for an initial phase
calibration (restricting to baselines longer than 2~m to avoid ground
spill over).

The electronic phase stability of the CBI system is better than 10 deg
over the timescales between primary phase calibrations. However the
pointing accuracy of the mount is approximately 0.5 arcmin
root-mean-square (rms).  Due to the co-mounted nature of the CBI
antennas this appears as correlated phase errors in the visibilities
and translates to an equivalent pointing uncertainty in the map.  The
pointing error can be measured, and removed, by interleaved
observations of a nearby phase calibrator. For the CBI observations on
2004 Dec 20, the calibrator J0541-051 was observed, and the resulting
pointing error correction was 62 arcsec.  For the CBI2 observations no
sufficiently bright calibrator was observed. We note that these
pointing errors and corrections are in any case a small fraction of
the synthesised beam size (6 arcmin for CBI and 4.5 arcmin for CBI2)
and do not significantly affect the positional matching of features in
the CBI maps with data at other wavelengths described below
\citep[see][for tests on the pointing accuracy of a similar
  dataset]{cas06}.


%

\begin{figure*}
\includegraphics[width=\textwidth,height=!]{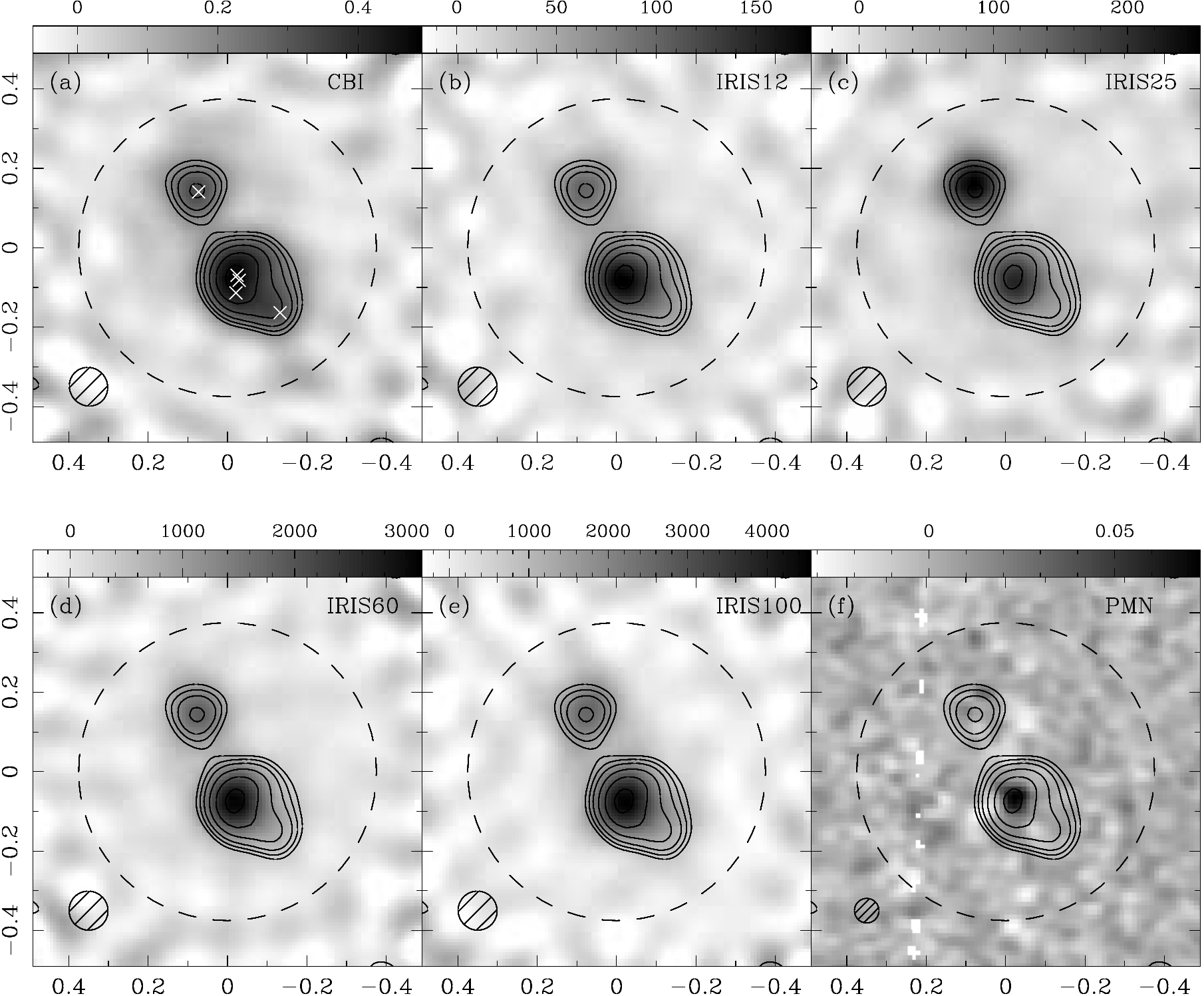}  
\caption{ Comparison of CBI contours and various templates. The units
  are Jy~beam$^{-1}$. The dashed circle corresponds to the CBI primary
  beam, the small circle outlined is the CBI synthesized beam. (a):
  restored CBI image. Contours are taken at 95, 80, 65, 52, 43 and
  35\% of the peak emission. White crosses indicate the brightest
  stars: from north to south, HDE290861, HD38563N, HD38563S and
  HD38563C. (b), (c), (d) and (e): restorations of CBI-simulated
  visibilities on the IRIS images at 12, 25, 60 and 100~$\mu$m, with
  contours from CBI. (f): PMN image at 4.85~GHz with CBI
  contours.\label{figure:corr1} Coordinates follow from
  Fig.~\ref{figure:large}.  }
\end{figure*}

\begin{figure*}
\includegraphics[width=\textwidth,height=!]{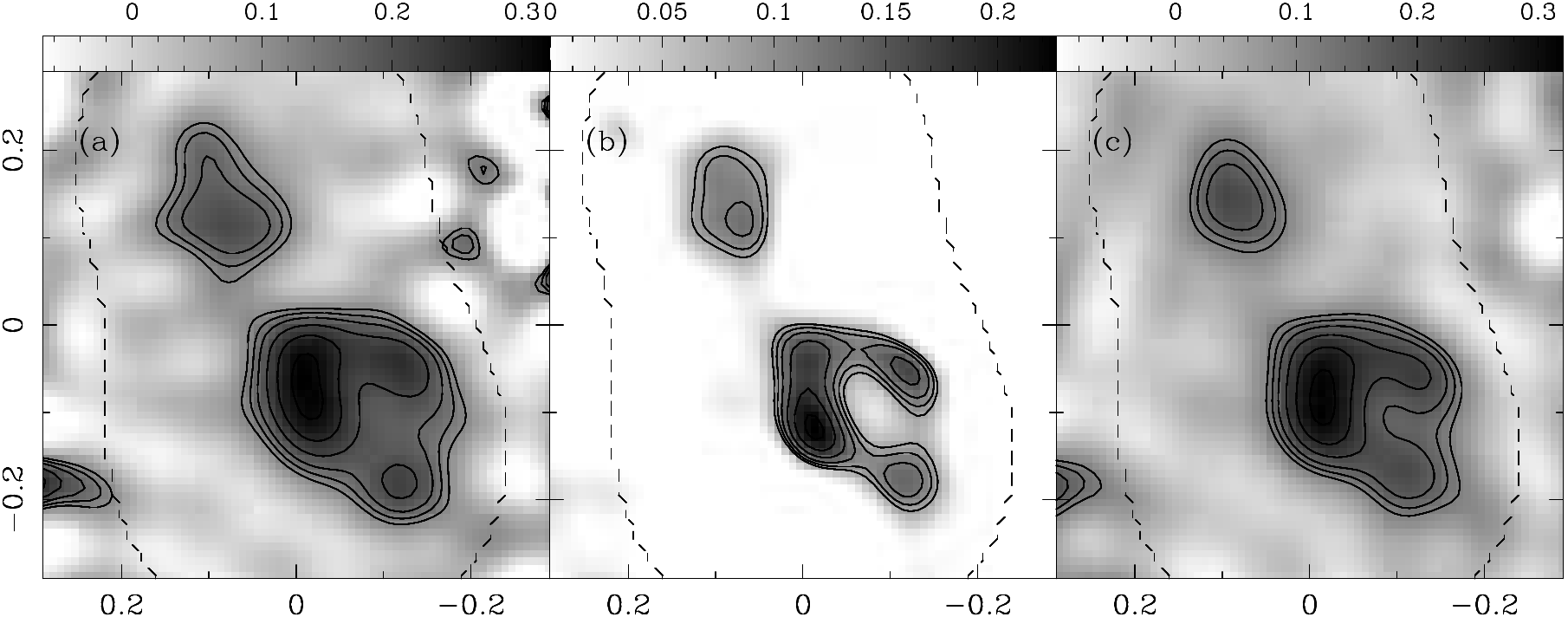}  
\caption{ Comparison of model-fit and MEM reconstructions of CBI2 data
  for M~78. Contours were taken at 95, 80, 65, 52, 43 and 35\% of the
  peak emission for each image. The dashed line corresponds to a noise level
  of twice the minimum noise. (a) Model-fitted
  reconstruction with contours. The maximum in the image is
  0.32~Jy~beam$^{-1}$. (b) MEM
  model with contours. The maximum is of 0.23~MJy~sr$^{-1}$. (c) Restored MEM
  image, the model was convolved
  with the clean beam and residuals were added, the peak intensity in this
  image corresponds to 0.32~Jy~beam$^{-1}$. All images are
  corrected for the primary-beam (including the MEM model). Coordinates
  follow from Fig.~\ref{figure:large}. \label{figure:cbi2}}
\end{figure*}

We fitted the CBI visibility data with 3 elliptical Gaussians,
corresponding to NGC~2068, NGC~2071 and NGC~2064. The restored image
is shown in Fig.~\ref{figure:corr1}a. In the case of CBI2 the
components are coincident with the four nebulae NGC~2068, NGC~2071,
NGC~2067 and NGC~2064. The Gaussian model convolved with the
synthesized beam plus residuals of the CBI and CBI2 data are shown in
Fig.~\ref{figure:corr1}a and Fig.~\ref{figure:cbi2}a,
respectively. Hereafter we refer to these images as ``model-fitted''
images. The CBI2 image shows clearly the improved resolution of CBI2,
and the increased flux loss of CBI2 due to reduced coverage at short
spacings.

For CBI2 we also implemented the MEM algorithm \citep[as
  in][]{cas06}. The MEM reconstruction provides a model sky consistent
with the data, prior to convolution with the synthetic beam. We show a
comparison of the two models in Fig.~\ref{figure:cbi2}.

The expected thermal noise for the CBI image is 12.78 mJy~beam$^{-1}$,
for CBI2 we have two values, one for M78 of 9.94 mJy~beam$^{-1}$ and
another for NGC2071, of 11.91 mJy~beam$^{-1}$. The achieved noise can
be measured by taking the rms dispersion of a region within the
half-power contour of the mosaiced primary beams, before correcting
for it. The noise in Fig.~\ref{figure:corr1}a, for CBI1, is
15~mJy~beam$^{-1}$, and the noise in Fig.~\ref{figure:cbi2}a, for
CBI2, is 20-30 mJy~beam$^{-1}$ (the combination of two fields results
in a varying noise map, we take 30 mJy~beam$^{-1}$ as a conservative
value).

\subsection{Auxiliary Data}

The images that are used to make a morphological comparison were taken
from IRIS \citep[][]{miv05}, a re-processing of the {\em IRAS} survey
\citep{whe91} at 12, 25, 60 and 100~$\mu$m. We use these infrared data
to test for correlation between the 31~GHz emission and the 
emission from interstellar dust. The mid-IR emission at 12~$\mu$m is
thought to be produced by stochastic heating of very small grains
(VSGs; e.g. \citet{dra01}). The heat capacity of these grains is small
enough so that the  absorption of a single UV photon will increase its
temperature to above 100 K, but their small cross section makes it
unlikely that they encounter many simultaneously. Thus VSGs are not in
equilibrium with the radiation field. At longer IR wavelengths, (such
as in the 60 and 100~$\mu$m IRIS bands), the emission is due to
classical grains at $T \sim 20 - 50$ K.

%

The resolution of the IRIS data is not fine enough to allow a
comparison with CBI2\footnote{as the IRIS resolution, of 4~arcmin,
  approaches that of CBI2, the simulation of CBI2 observations on the
  IRIS images will yield coarser resolutions than real CBI2 maps}. To
trace small-scale structure, that could be compared with the improved
resolution of CBI2, we used archival data acquired with the IRAC
instrument aboard the {\em Spitzer} Space Telescope. This image is
taken at 8~$\mu$m and traces VSGs and PAHs, like IRIS at
12~$\mu$m, but with a finer resolution of 4$\arcsec$ that allows
proper point-source subtraction and comparison with CBI2 data. The
IRAC image is shown in Fig.~\ref{fig:tript}a. 
 
%


Point sources from IRAC were subtracted using a median filter. For the
remaining sources, they were subtracted individually by fitting a
Gaussian to the source and then subtracting it. CBI observations were
simulated on both the IRIS and IRAC images using the MOCKCBI package
(Pearson 2000, private communication), in order to have equivalent
visibility sampling when comparing with CBI and CBI2 images.  This
will remove any uniform background from IRIS and IRAC. When simulating
IRAC data we embedded the image in a background taken from IRIS
12~$\mu$m, to extend the field of the IRAC 8~$\mu$m mosaic to at least
3 times the size of the primary beam. When embedding IRAC~8~$\mu$m
in IRIS 12~$\mu$m we fitted a plane to the difference of emission
from the border of the images in order to remove most of the
discontinuities and to remove the contribution from zodiacal light in
the 8~$\mu$m image (IRIS images have only residual contribution
from zodiacal light). The simulation is also useful to acquire
visibility-plane data, which we used to make linear correlations to
obtain a quantitative comparison between dust and 31~GHz emission. The
simulations obtained will hereafter be referred to as CBI-simulated or
CBI2-simulated images.


We also carried out a morphological comparison of the CBI2 data with
PMN (at 4.85 GHz) \citep[][with a resolution of 4.1$\arcmin$, see
  Sec.~\ref{sec:ffintens}]{con91}, NVSS (1.4~GHz) \citep[][with a
  resolution of 45$\arcsec$]{con98} and SHASSA \citep[at 6563\AA, and
  with a resolution of 48$\arcsec$,][]{gau01}. These templates trace
free-free emission, which is present, at some level, at 30~GHz. It is
necessary to quantify the intensity and spatial distribution of the
free-free component in order to test for the existence of an excess at
30~GHz. These comparisons are discussed in Sec.~\ref{sec:ffintens}.

For measurements of flux density we have used images taken from {\em
  WMAP} which give information about the continuum radiation on larger
scales. The {\em WMAP} channels are at 22.8, 33.0, 40.7, 60.7 and 93.5
GHz, with resolutions of 0.88, 0.66, 0.51, 0.35 and 0.22 degree
respectively \citep{lim09}. We also used data from the Rhodes/HartRAO
survey at 2.326~GHz with a resolution of 20$\arcmin$ \citep{jon98}.

\subsection{Flux Loss Correction and  Spectral Energy
  Distribution} \label{sec:obs_sed}

The CBI, being an interferometer with no total power measurements, is
insensitive to an uniform background. The lack of total power also
removes part of the extended emission, thus high-pass filtering the
sky intensity field.

In order to estimate the level of flux loss, due to missing short
spacings, we used the IRIS images as templates, shown in
Fig.~\ref{figure:corr1} to correlate with the CBI signal. We calculated
the integrated fluxes in a 45$\arcmin$ aperture from both raw and
CBI-simulated data at each of the IRIS wavelengths, and compared
them. This showed an average flux loss of $34\pm8$\% among the IRIS
bands. This means that most of the flux in the IRIS images is within
the compact objects corresponding to NGC~2071, NGC~2068 and NGC~2064.

Table~\ref{table:sed} summarises the SED of M~78, for flux densities
extracted in a circular aperture 45$\arcmin$ in diameter, corresponding
to the CBI primary beam diameter, and centered on RA: $05^h46^m46^s.7$
and DEC: $+00\degr 00\arcmin 50\arcsec$. Our choice of aperture
contains the bulk of the M~78 region, and matches the signal seen by
CBI.  Using the same aperture diameter we also calculate the flux
density in 8 reference fields which are marked as crosses in
Fig.\ref{figure:large}. Subtracting the flux density in each reference
field from the flux density centered on M~78, we remove the uniform
background, and estimate the error in this background level with the
root-mean-square (rms) dispersion of the 8 different
background-subtracted flux densities.

\begin{table}
\centering
\caption{\label{table:sed} Flux densities extracted in a circular aperture
  45$\arcmin$ in diameter, and centred on M~78.}
\begin{tabular}{cccrrrrrrrr}
\hline
$\nu^a$ & $F_\nu^b$ & $F_\nu^c$ & \\
  2.326  &  $7.08\pm0.21$ &  $0.89\pm0.48$         \\ 
 22.8 &   $4.40\pm 0.31$ &  $2.34\pm0.66$  \\ 
 33.0 &   $4.34\pm 0.47$ &  $2.79\pm0.81$  \\
 40.7 &   $4.37\pm 0.59$ &  $3.18\pm0.98$  \\
 60.7 &   $5.07\pm 0.69$ &  $4.68\pm1.44$  \\
 93.5 &   $20.20\pm 1.47$ &  $19.55\pm3.60$      \\
 31   &   $3.96\pm 0.30$$^d$ &  $2.73\pm0.28$$^d$\\
 3000 &   $32418\pm3220$ &  $29700\pm3310$ \\ \hline
\end{tabular}
\begin{flushleft}
$^{a}${Frequency of observation in GHz.}
$^{b}${Flux density in Jy.}
$^{c}${Flux density in Jy after differencing with the
  reference fields.}
$^{d}${This value is obtained from IRIS 100~$\mu$m-31~GHz
  correlated emission, using the slope found from the visibility
  correlation.}
$^{e}${Origin of the data, for each frequency (in GHz): 2.326, \citet{jon98};
  22.8, 33.0, 40.7, 60.7, 93.5, \citet{lim09}; 31, this work; 3000,
    \citet{miv05}}
\end{flushleft}
\end{table}


The error bars in Table~\ref{table:sed} give 1~$\sigma$, with $\sigma
= \sqrt{\sigma^2_\mathrm{noise}+\sigma^2_\mathrm{diff}}$, where:
\begin{itemize}
\item $\sigma_\mathrm{diff}$ is the rms dispersion of the results from
  differentiations against the different reference fields, 
\item $\sigma_\mathrm{noise} = \Delta\Omega I_\mathrm{rms}
  \sqrt{N_\mathrm{beam}~N_\mathrm{apert}}$, where $\Delta\Omega$ is
  the solid angle of one pixel, $ I_\mathrm{rms}$ is the nominal
  thermal noise in specific intensity, when available, or  the
  root-mean-square dispersion of a $\sim$0.5~deg square box devoid of
  signal, $N_\mathrm{beam}$ is the number of pixels in a synthesized
  beam for each image, $N_\mathrm{apert}$ is the number of pixels in the
aperture, in this case a diameter of 45$\arcmin$. 
\end{itemize}
An additional term $\sigma_\mathrm{cal}$ could have been added in
quadrature to the 1~$\sigma$ errors on a flux density $F$,
$\sigma_\mathrm{cal} = 0.1~F$, as a representation of systematic errors
in calibration. However $\sigma_\mathrm{diff}$ dominates the error
budget, and $\sigma_\mathrm{cal}$ is negligible when added in
quadrature.

\section{Discussion} \label{sec:disc}

To characterize the emission from M~78 that may not be explained as
free-free and/or vibrational dust, i.e. the Rayleigh-Jeans tail of
sub-mm dust emission, we will perform a morphological and spectral
analysis. First, we will summarize a comparison in the sky plane with
various templates that trace different phases of dust and gas, to find
qualitatively which corresponds better with the emission found at
31~GHz. Secondly, we will correlate the CBI data with the CBI-simulated
templates, both in the visibility and sky planes. The visibility
cross-correlations give information on the average properties over the
CBI primary beam. Thirdly, another diagnostic is fitting a SED. This
will give us the spectrum of {M 78}, which allows us to
estimate free-free and vibrational dust contributions at 31~GHz, and
infer the level of any excess.

\subsection{Qualitative  CBI - IR comparisons}  \label{sec:qual_CBI_IR}

In Fig.~\ref{figure:corr1}b-e we compare the CBI and IRIS
images. There is a qualitative correlation between CBI and IRIS, with
some variations. At 25~$\mu$m NGC~2071 is more prominent than
NGC~2068, which contrasts with IRIS 12, 60 and 100~$\mu$m. For the
other IRIS images it is difficult to tell by visual inspection which
correlates best with CBI.

\begin{figure*}
\begin{center}
\includegraphics[width=\textwidth,height=!]{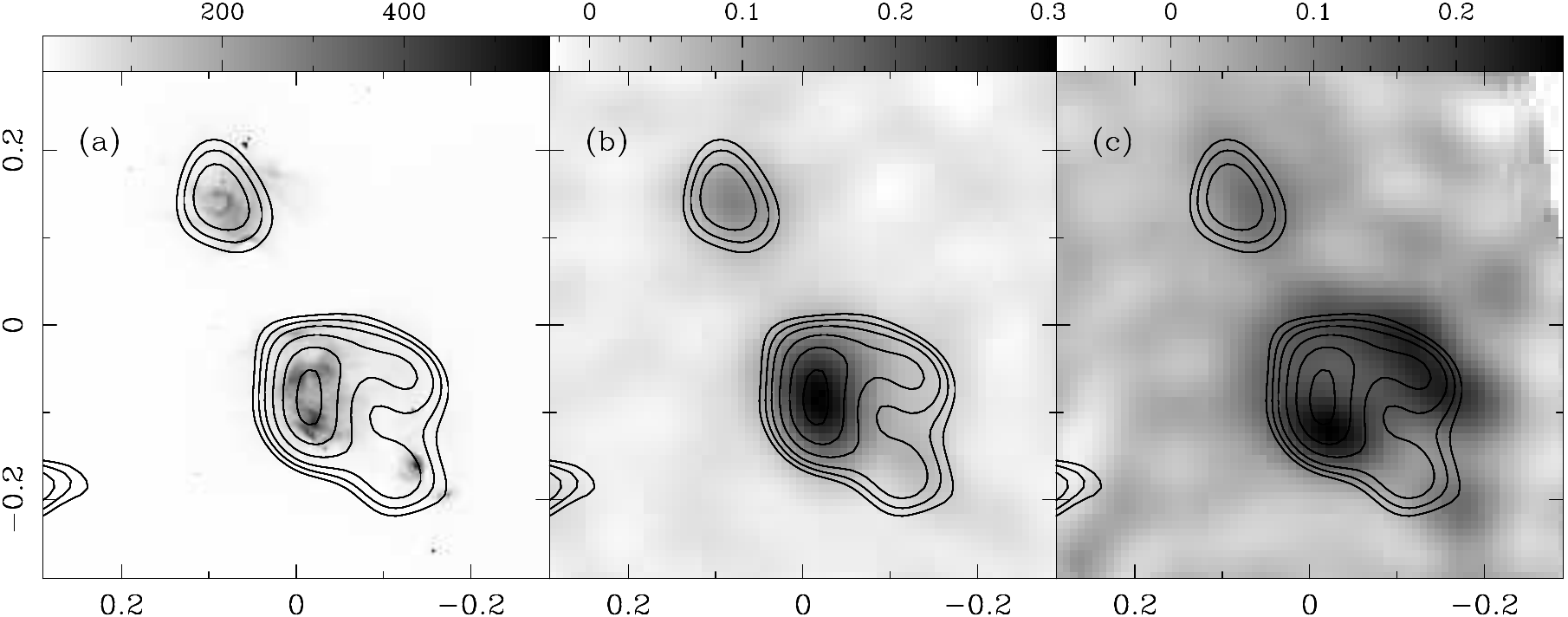}  
\end{center}
\caption{(a) Comparison between IRAC 8~$\mu$m (in grey scale) and
  the CBI2 MEM restored image (in contours). The intensities are in MJy~sr$^{-1}$ (b)
  In gray scale we have the  CBI2-simulated IRAC~8~$\mu$m image, scaled by the
  correlation slope with CBI2, against the restored image from CBI2, both
  reconstructed using model-fitting. (c) As in (b), but for simulations of CBI2
  observations on the ratio of IRAC~8~$\mu$m and the UV field
  $G_\circ$. Coordinates follow
  from Fig.~\ref{figure:large}. \label{fig:tript}}
\end{figure*}

In Fig.~\ref{fig:tript} we compare the CBI2 MEM model in contours with
the IRAC~8~$\mu$m image in grey scale. This shows us that the spatial
correlation obtained between IRIS 12~$\mu$m and CBI still holds on
smaller scales. We also see a radio peak with no IR counterpart towards
NGC~2067. In IRAC this nebula appears as diffuse emission, while in
CBI2 this peak is more intense than the one observed towards NGC~2064.
In Fig.~\ref{fig:tript}b we show the reconstructed CBI2-simulated
image from IRAC obtained by model-fitting compared with the CBI2
restored MEM image (Fig.~\ref{figure:cbi2}). We scaled the CBI2-simulated
visibilities by the cross-correlation slopes obtained in
Sec.~\ref{sec:CBI_IRcorr} (see Table~\ref{table:corr2}) to bring them to the
same range as the CBI2 data, and averaged 90 reconstructions, corresponding to
different realisation of Gaussian noise as given by the CBI2 visibility
weights. This noise-simulation process applied to the CBI2-simulated
IRAC visibilities, and subsequent averaging, avoids differences in
convergence with the models fit to CBI2, as could arise from the
varying visibility weights.

We see in Fig.~\ref{fig:tript} that emission from NGC~2064 is at
$24\pm 1$\% of that found for NGC~2068 in the CBI2-simulated 8~$\mu$m
image, while in CBI2 this percentage is $70\pm 7$\%. We also note that
the radio emission peaks are not coincident with illuminating stars,
particularly in NGC~2064 where the radio nebula is offset to the
southeast of the infrared counterpart. If the emission is due to
spinning dust, and if the spinning dust emissivities are fairly
independent of the radiation field \citep[e.g.][]{ali09,yv09}, we
expect the ratio of mid-IR and CBI images to be proportional to the
averaged UV intensity. So we do not expect a direct correspondence
between the radio and the IR. For instance, in the vicinity of the
brightest stars the mid-IR intensities will be enhanced relative to
the spinning dust intensities. The UV field will also affect the grain
size distribution, directly impacting the spinning dust
spectrum. Mid-IR specific intensities from dust grains are expected to
be proportional to the UV field, as well as to very small grain column
densities. This may explain the fact that NGC~2067 is fainter at
8~$\mu$m than at 31~GHz because this nebula does not contain stars
that are bright enough to have a strong UV field, but still has enough
dust column to radiate conspicuously in the radio.

We find that the morphological differences between IRAC~8~$\mu$m and
the 31~GHz continuum cannot be explained only in terms of the
proportionality of IR emission and the UV field.  Following the
procedure described in Sec.~\ref{sec:uvfield}, we have estimated the
UV field in M~78, parametrised as $G_\circ$. Fig.~\ref{fig:tript}c
presents the simulation of CBI2 observations on the ratio of
IRAC~8~$\mu$m to $G_\circ$. It is apparent from Fig.~\ref{fig:tript}c
that the radio/IR correlation worsens in M~78 after division by
$G_\circ$. The peak in NGC~2068 is shifted to the south when
compared to the observed from the UV field uncorrected IRAC image and the one
observed with CBI2. We also observe that emission corresponding to NGC~2067
appears clearly in Fig.~\ref{fig:tript}c, while in Fig.~\ref{fig:tript}a and b
it is missing.

%
%

\subsection{CBI - IR correlations} \label{sec:CBI_IRcorr}

For a quantitative analysis we performed linear correlations between
CBI and IRIS data and between CBI2 and IRAC at 8~$\mu$m, both in
the visibility and sky planes. We used different diagnostics in each
case. On the visibility plane we used reduced $\chi^2$ and the linear
correlation coefficient $r$, which we calculated using:
\begin{equation}
r = {N~\sum{(V_{31}~V_t)}-\sum{(V_{31})}~\sum{(V_t)}\over{\sqrt{N~\sum{(V_{31}^2)}-(\sum{V_{31}})^2}~\sqrt{N~\sum{(V_t^2)}-(\sum{V_t})^2}}},
\end{equation}
where $N$ is the number of visibilities, $V_{31}$ are the visibilities at
31~GHz, $V_t$ are the visibilities from the simulated template and the
sums extend over every visibility. This last coefficient will be referred to
as $r_\mathrm{vis}$.  

The sky plane correlations were carried out within circular areas,
with radii of 45~arcmin (i.e. with diameters twice the CBI primary
beam FWHM), and centred on each region of interest (i.e. M~78 for CBI1
and NGC~2068, NGC~2071 for CBI2). These sky plane correlations were made with
modelfitted images. For these correlations we also use
the $r$ coefficient, changing now visibilities for intensities and $N$
for the number of pixels. This coefficient will be referred to as
$r_\mathrm{sky}$. We also use the normalized cross-correlation
coefficient,
\begin{equation}
C = {\sum(I_{31}~I_t \alpha )\over{\sum(I_{31}^2)}},
\end{equation} 
where $I_{31}$ and $I_t$ are the intensities at 31~GHz and from the
template respectively and $\alpha$ is defined by $I_{31} = \alpha I_t$
and determined by linear regression \citep[this definition of $C$ is
  as used in][except there is a typo in their Eq.~1, where $\alpha$ is
  missing]{cas06}. Here again, the sums are taken over all pixels
falling within a circular aperture 90~arcmin in diameter. This choice
of aperture is meant to include all of the signal seen by CBI.

To obtain statistics on $r_\mathrm{sky}$ and $C$ we averaged 1000
values of $r_\mathrm{sky}$ and $C$ taken on different reconstructions
of the CBI data, each with the addition of different realisations of
Gaussian noise to the visibilities (given by the visibility
weights). The values shown in Table~\ref{table:corr} and
Table~\ref{table:corr2} correspond to the mean value of a normal
distribution fitted to the data so obtained, and the errors correspond
to 1~$\sigma$.

\begin{table}
\caption{\label{table:corr}Linear Correlations Results for CBI versus
  IRIS, both for visibility and sky planes.}
\begin{tabular}{cccccrrrr}
\hline
Template & $\chi^2/{\nu}^a$ & $r_\mathrm{vis}^b$ &$C^c$ &
$r_\mathrm{sky}^d$  \\ 
12 $\mu$m  & 1.40 & 0.776 & $0.683\pm0.024$ & $0.786\pm0.019$ \\
 12 $\mu$m$^{e}$  & 1.39 & 0.779 & $0.774\pm0.025$ & $0.786\pm0.018$ \\
 25 $\mu$m  & 1.59 & 0.768 & $0.644\pm0.023$ & $0.739\pm0.021$ \\ 
 25 $\mu$m$^{e}$  & 2.15 & 0.719 & $0.627\pm0.022$ & $0.717\pm0.017$ \\
 60 $\mu$m  & 1.58 & 0.755 & $0.651\pm0.020$ & $0.787\pm0.017$ \\
 60 $\mu$m$^{e}$  & 1.28 & 0.789 & $0.750\pm0.024$ & $0.819\pm0.018$ \\
 100 $\mu$m  & 1.39 & 0.777 & $0.699\pm0.021$ & $0.802\pm0.019$ \\ 
 100 $\mu$m$^{e}$  & 1.58 & 0.754 & $0.761\pm0.027$ & $0.761\pm0.018$
 \\ \hline
\end{tabular}
\begin{flushleft}
$^{a~}${Reduced $\chi^2$, taken in the visibility plane.\\
  $\nu$ is the number of degrees of freedom, i.e. twice the number of
  visibilities (counting real and imaginary parts)}\\
$^{b~}${{Linear} correlation coefficient from the visibility
  correlations .} \\
$^{c~}${Normalized cross-correlation, from the sky plane correlation.}\\
$^{d~}${{Linear} correlation coefficient, obtained from the sky plane
  correlations.}\\
$^{e~}${Comparisons made with UV field corrected IRIS images.} \\
\end{flushleft}
\end{table}

\begin{table}
\caption{\label{table:corr2}Linear Correlations Results for CBI2 versus
  IRAC 8~$\mu$m}
\begin{tabular}{cccccrrrr} \hline
{Template} & {$\chi^2/{\nu}$ $^{a}$} &
{$r_\mathrm{vis}$$^{b}$} & {$C$$^{c}$} &
{$r_\mathrm{sky}$$^{d}$} \\
 NGC~2068  &  3.44 &  0.254 & $0.473\pm0.017$ & $0.739\pm0.020$ \\ 
 NGC~2068$^{e}$  &  3.92 &  0.129 & $0.409\pm0.015$ & $0.677\pm0.019$ \\ 
 NGC~2071  &  3.99 &  0.205 & $0.347\pm0.028$ & $0.530\pm0.034$ \\
 NGC~2071$^{e}$  &  4.08 &  0.180 & $0.402\pm0.031$ & $0.516\pm0.036$
 \\ \hline 
\end{tabular}
\begin{flushleft}
$^{a}${Reduced $\chi^2$ of the  
  visibility correlation, where  $\nu$ is the number of  degrees of freedom}
$^{b}${{Linear}  correlation coefficient, obtained from the visibility correlation.}
$^{c}${Normalized cross-correlation, from the sky plane correlation.}
$^{d}${{Linear} correlation coefficient, obtained from the sky plane
  correlations.}
$^{e}${Comparisons made with UV field corrected IRAC
  images.}
\end{flushleft}
\end{table}

Table~\ref{table:corr} summarizes numerical results from the
correlations between CBI and IRIS. As we can see the best results
independently of the diagnostic (reduced $\chi^2$, $C$,
$r_\mathrm{vis}$ or $r_\mathrm{sky}$) correspond to 12~$\mu$m and
100~$\mu$m, which both, given the uncertainties, correlate equally
well with the CBI data. The worst correlations are obtained at 25~$\mu$m,
except for the $r_\mathrm{vis}$ diagnostic, where it gives a value closer to one
than the obtained result at 60~$\mu$m.

For correlations with the finer resolution data of CBI2 we used IRAC
8~$\mu$m as a template corresponding roughly to IRIS
12~$\mu$m. However the two channels do not probe exactly the same
medium, we have different widths and line contributions. In the case
of 12~$\mu$m the channel ranges from $\lambda = 7~\mu$m to $\lambda
= 15.5~\mu$m ($\Delta \lambda = 8.5~\mu$m) and includes PAHs features
at 7.7 and 11.3~$\mu$m and H$_2$ lines at 8.026, 9.665 and
12.3~$\mu$m. For 8~$\mu$m the range is from $\lambda =
6.5~\mu$m to $\lambda = 9.5~\mu$m ($\Delta \lambda = 3~\micron$)
with contributions from the PAH feature at 7.7~$\mu$m and H$_2$
lines at 6.91 and 8.026~$\mu$m.


The correlation diagnostics worsen when testing with CBI2-simulated
data from IRAC~8~$\mu$m, as summarised in Table~\ref{table:corr2}.
The degradation of the correlation when comparing CBI2 and
IRAC~8~$\mu$m might be due to the intrinsic morphological differences
pointed out in Sec.~\ref{sec:qual_CBI_IR}, which stand out with the
finer resolution of CBI2 and IRAC, compared to the CBI and IRIS tests.

\subsection{Stellar UV field and VSG column density map} \label{sec:uvfield}

Under the spinning dust hypothesis we expect both the mid-IR and radio
emission to be proportional to the column density  of VSGs.  We assume that
mid-IR emission is due to stochastically heated VSGs. Thus the mid-IR
emission will also be proportional to the UV field. We also assume
that spinning dust is proportional to the column of VSGs, and is fairly
independent of the local UV field. 

\begin{figure}
\begin{center}
\includegraphics[width=\columnwidth,height=!]{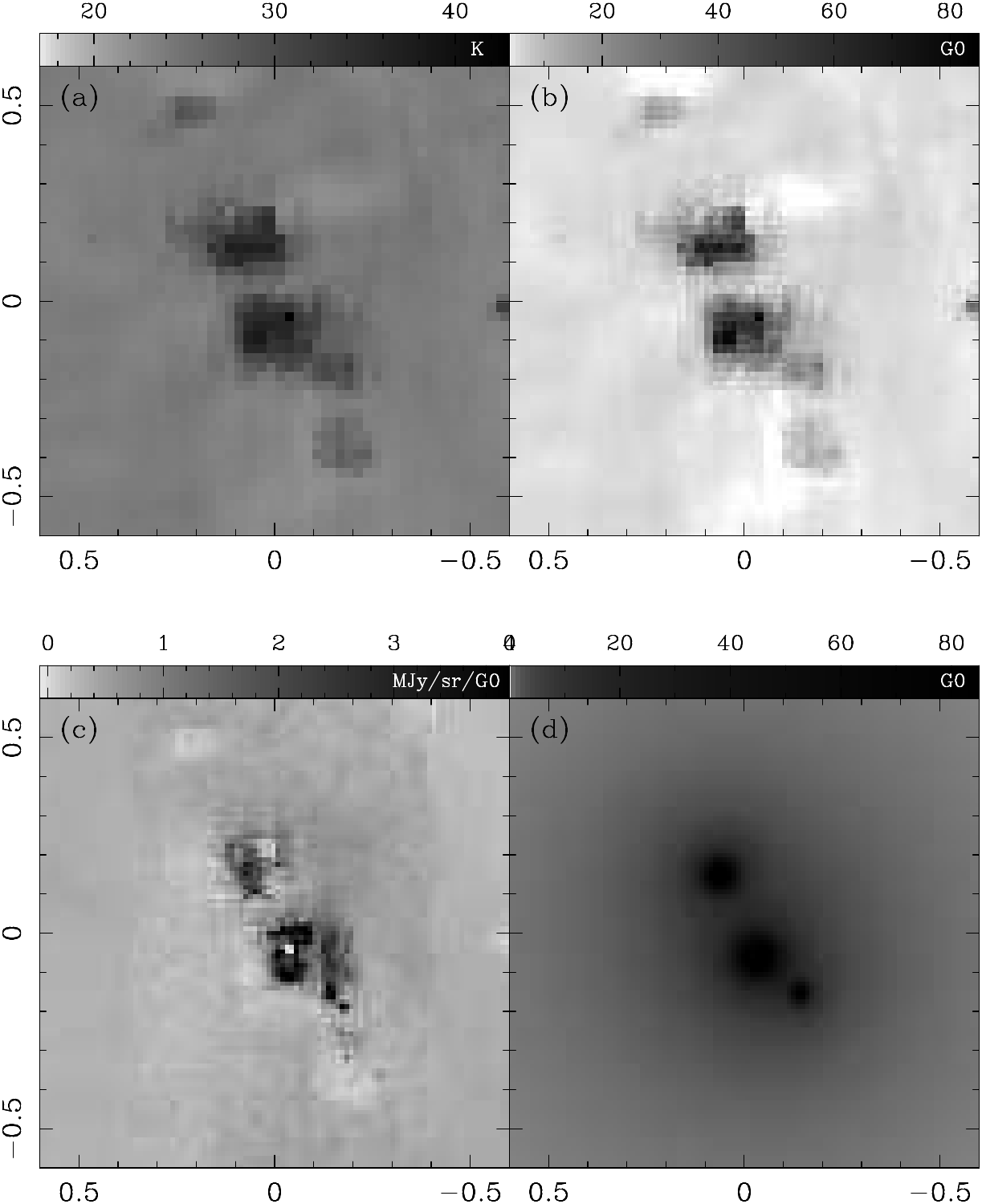}  
\end{center}
\caption{\label{fig:temperature} Proxies for the temperature and UV
  field in M~78. a) Dust temperature map obtained from IRIS~60/100, b)
  implied $G_\circ$ map with a dust emissivity $\beta = 2$. c)
  IRAC~8$\mu$m filtered to remove point sources (i.e. stars), and
  resampled at IRIS resolutions.  d) Expected $G_\circ$ map from the
  stellar content of M~78, and in the absence of extinction. The
  grey-scale has been chosen to match b); $G_\circ$ diverges as the
  inverse-squared projected distance to the UV sources. Coordinates follow
  from Fig.~\ref{figure:large}. }
\end{figure}

Then to obtain a more accurate template of the VSG column we must
divide the mid-IR images by an estimate of the UV field. We estimated
the dust temperature $T_d$ with the IRIS images at 60 and
100~$\mu$m. The resulting $T_d$ map shown on
Fig.~\ref{fig:temperature}a was obtained by fitting a modified
black-body to each pixel using an emissivity index for big grains of
$\beta = 2$ (representative of the emissivity index in the far-IR).
The $T_d$ map can be converted to an equivalent radiation field using
$G_\circ = (T_d/17.5~\mathrm{K})^{\beta+4}$ \citep[as
  in][]{ysa09}. The result is shown in Fig.~\ref{fig:temperature}b.

We also fitted a modified black-body to the integrated flux at 100 and
60~$\mu$m obtaining an average temperature of 32~K and a spectral index of
1.83, consistent with big grains. The spinning dust emission should arise from
the VSG population that is traced by mid-IR emission. Calculating the expected
emission from this modified black-body at 12~$\mu$m we found that it is
negligible ($\sim 10^{-5}$~Jy), while the value extracted from the 12~$\mu$m
image is $\sim 1200$~Jy. Thus we expect a significant population of VSGs. This
is also deduced from the 8~$\mu$m image which shows  significant emission from
the 7.6~$\mu$m PAH band.

The UV luminosities of HD38563N, HD38563S, HD38563C, HDE290861 and
SSCV 111 are extracted from the ATLAS12 model atmospheres
\citep{cast05}. The values we used for the stellar parameters are
listed on Table~\ref{table:strom}. In connection with the discussion
on C\,{\sc i} continuum in Sec.~\ref{sec:CIcontinuum}, we include the
Str\"omgren radius of the C\,{\sc ii} region around each star using
values of $n_e = 1$ cm$^{-3}$ and $T_e = 50$ K, and recombination
coefficients from \citet{nah97}.

\begin{table*}
\caption{\label{table:strom} Parameters for the brightest stars embedded within M~78.}
\begin{tabular}{cccccc}\hline
{Name} & {Spectral Type} & {Temperature$^{a}$} &
{Surface Gravity$^{b}$} & {Luminosity$^{c}$} & {C II
  Str\"{o}mgren Radius$^{d}$}\\
 HD38563N & B2\,{\sc iii} & 20000 & -1 & $1.6~10^4$ & 6.06 \\
 HDE290861 & B2\,{\sc v} & 22000 & -0.5 & $6.6~10^3$ & 4.84 \\
 HD38563S & B5\,{\sc v} & 15000 & -0.5 & $6.9~10^2$ & 1.39 \\
 HD38563C & A0\,{\sc ii} & 10000 & -1.5 & $1.3~10^3$ & 0.11 \\
 SSCV 111 & B3\,{\sc v} & 19000 & -0.5 & $2.7~10^3$ & 3.16 \\ \hline
\end{tabular}
\begin{flushleft}
$^{a}${In K.}\\
$^{b}${In $\log(g/g_\odot)$.}\\
$^{c}${In $L_\odot$.}\\
$^{d}${In pc using $n_e = n_{\mathrm{C}^+} = 1$~cm$^{-3}$.}\\
\end{flushleft}
\end{table*}

Knowledge of the stellar content can be used to estimate the
circumstellar UV field, by diluting the stellar UV luminosities
according to the inverse-square of the projected distance.  This rough
estimate of the $G_\circ$ map, shown in Fig.~\ref{fig:temperature}d,
assumes that the layers of M~78 exposed to the stellar UV field do not
overshadow. In other words, Fig.~\ref{fig:temperature}d assumes that
the mid-IR emission stems from regions directly exposed to the
radiation from the illuminating stars, without intervening
extinction. The similarity of the maps in Fig.~\ref{fig:temperature}b
and Fig.~\ref{fig:temperature}d helps as a sanity check on
Fig.~\ref{fig:temperature}b.

With an estimate of the UV field in M~78, as parametrised by $G\circ$,
we can divide the IRAC~8$\mu$m image by the $G_\circ$ map to produce a
VSG column density map. This estimative column density map, in
arbitrary units, is shown on Fig.~\ref{fig:temperature}c.  The spike
near the origin of coordinates in Fig.~\ref{fig:temperature}c
corresponds to HD38563N.

\subsection{CBI and CBI2  correlations with the VSG column density map}

\begin{figure}
\begin{center}
\includegraphics[width=\columnwidth,height=!]{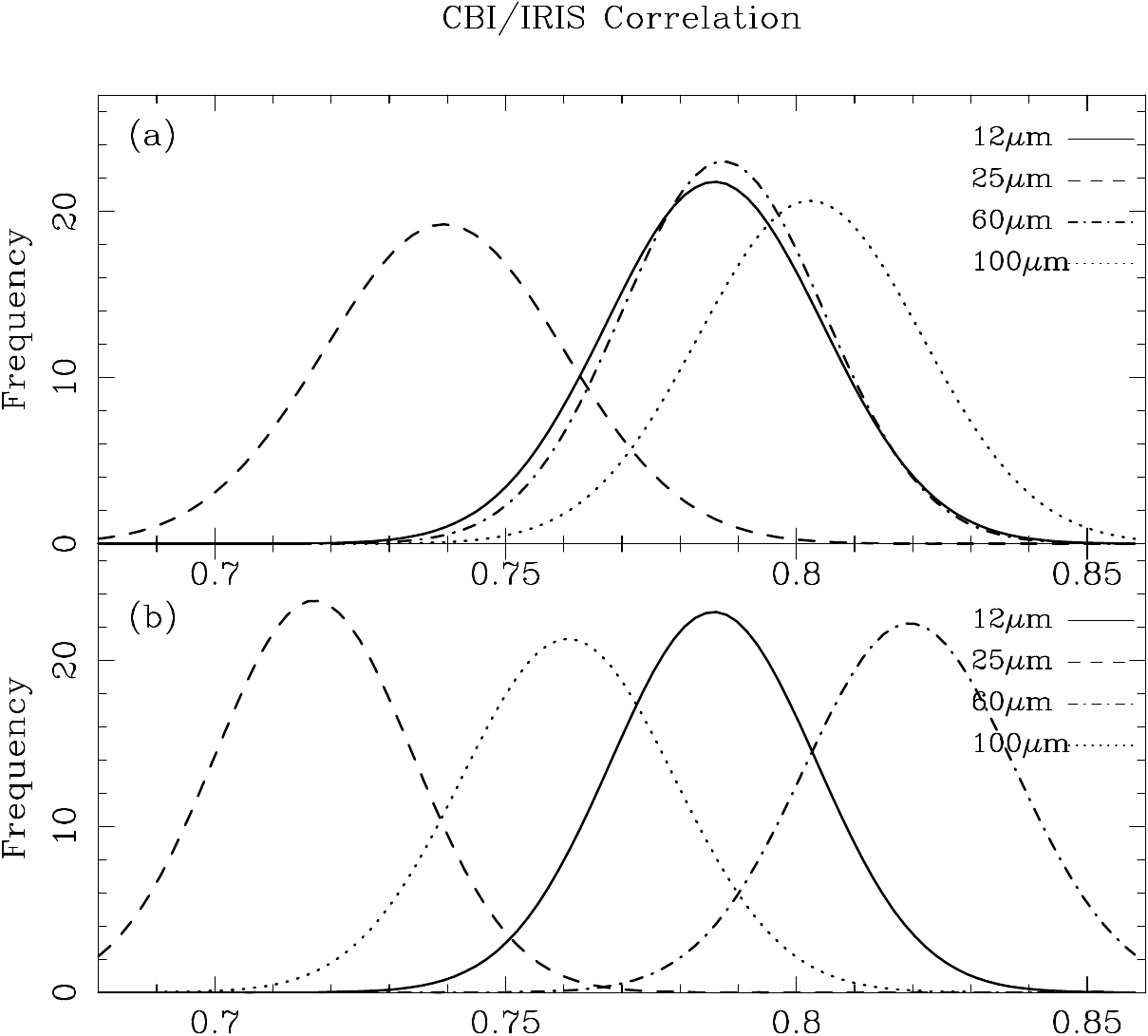}  
\end{center}
\caption{(a) Fitted distributions to Monte-Carlo noise simulation of the
  correlation coefficient $r_\mathrm{sky}$ between CBI and IRIS for M~78. The
  solid lines corresponds to the 12~$\mu$m templates, the dashed lines to
  25~$\mu$m, the dash-dotted lines to 60~$\mu$m and the dotted lines to
  100~$\mu$m. The Gaussian parameters are listed in
  Table~\ref{table:corr}. (b) Distributions obtained from UV-field corrected
  templates. The corresponding parameters are also listed in
  Table~\ref{table:corr}. \label{figure:corriris}}
\end{figure}

In Fig.~\ref{figure:corriris}a we see the
gaussian distributions fitted to the histograms of the
$r_\mathrm{sky}$ coefficient for the different IRIS bands before performing
the UV field correction. The two
gaussians fitted to the histograms at 12 and 60~$\mu$m are almost
coincident while at 100~$\mu$m we find the best correlation,
$r_\mathrm{sky} = 0.802\pm0.019$. Finally at 25~$\mu$m we find the
worst correlation result, as was expected because this is the only
template where NGC~2071 appears brighter than NGC~2068.

In Table~\ref{table:corr} we give the results for IRIS-CBI
correlations, corrected by our estimate of the UV field. We see that
the best correlation now corresponds to 60~$\mu$m, which improves from
the uncorrected version, as reflected in the 3~$\sigma$ increase in
$r_\mathrm{vis}$, $C$ and $r_\mathrm{sky}$.  For 12~$\mu$m the
improvement is slight.  In general we see that the values found for 25
and 100~$\mu$m suffer a worsening when compared with the UV-field
uncorrected results.  In the case of 25~$\mu$m, we see again the worst
correlation results, as expected for here the brightest nebula is
NGC~2071 and not NGC~2068. The Gaussian fit to the distribution of the
$r_\mathrm{sky}$ coefficient is shown in Fig.~\ref{figure:corriris},
where we can compare cross-correlations with templates corrected and
uncorrected by the UV field.

\begin{figure}
\begin{center}
\includegraphics[width=\columnwidth,height=!]{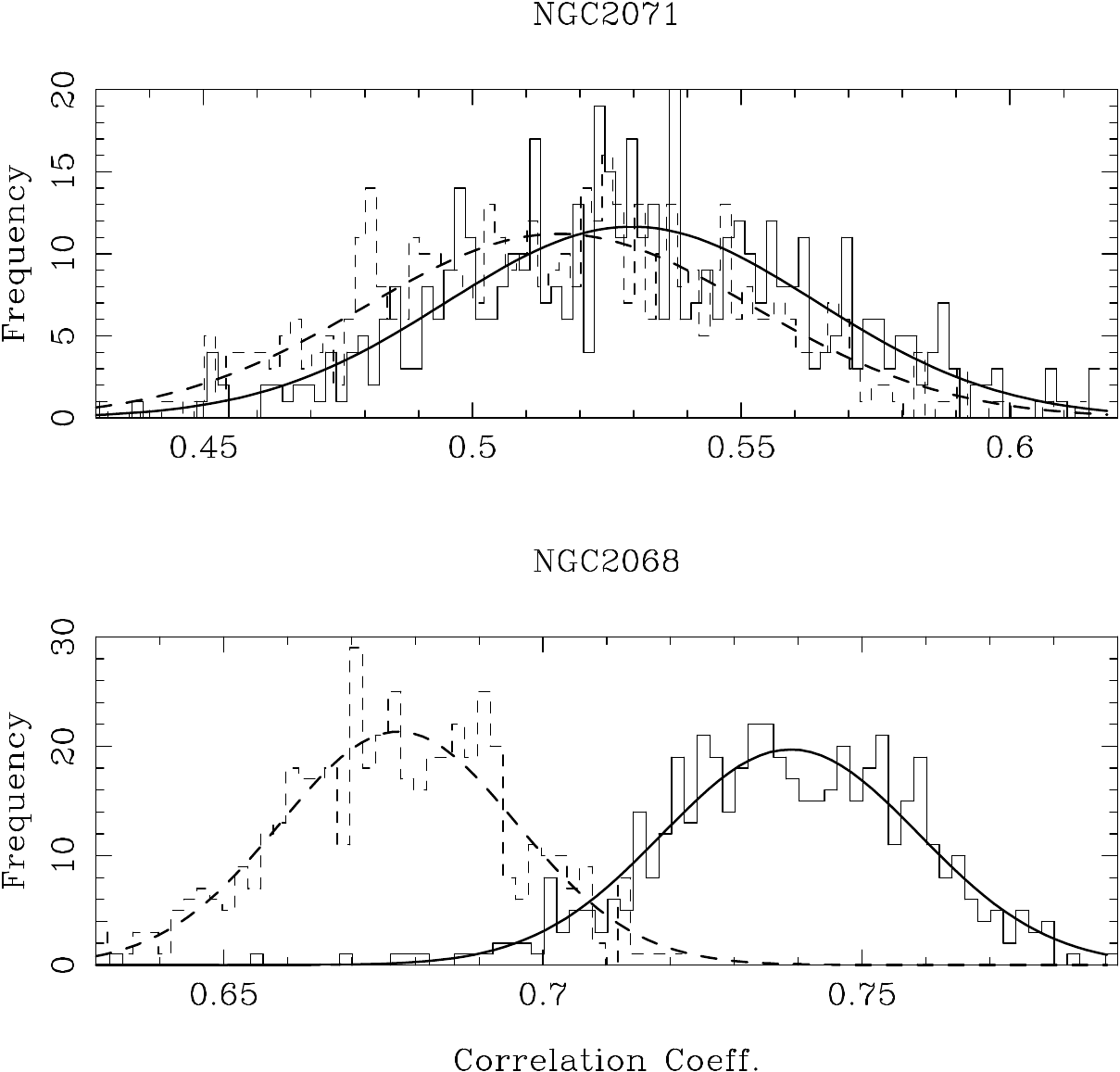}  
\end{center}
\caption{Distribution obtained from Monte-Carlo noise simulations of
  the correlation coefficient $r_\mathrm{sky}$ between CBI2 and IRAC
  for NGC~2068 and NGC~2071. The solid line histogram corresponds to
  the data obtained without the UV field correlation, while the dashed
  line one is the corrected distribution. We also show gaussian fits
  for each case, whose parameters are listed in
  Table~\ref{table:corr2}.\label{figure:distirac}}
\end{figure}

The UV field corrected results for IRAC-CBI2 correlations are given in
Table~\ref{table:corr2}. In Fig.~\ref{figure:distirac} we show the
distribution of the correlation coefficient $r_\mathrm{sky}$ obtained
for each nebula, with and without the UV field correction. In NGC~2071
the correlation indicators are generally slightly lower after division
by $G_\circ$, but the decrement is within the noise. In NGC~2068 all
correlation indicators decrease markedly, with significant decrements
in $C$ and $r_\mathrm{sky}$ over 3~$\sigma$.

%
%

The estimates of the VSG column density maps produced by dividing the
IR images by the UV-field should be taken with some caution.  Apart
from the uncertainties in the UV-field estimates, there is also the
possibility that the grain population may be influenced by the UV
field, in particular in its size distribution. To make a more complete
comparison is beyond the scope of this paper.


\subsection{Expected C\,{\sc i} continuum} \label{sec:CIcontinuum}

Estimates of the C\,{\sc i} continuum level can be made under the
assumption of ionisation balance for the stellar content of M~78 with
$T_e$ ranging from 20 to 50~K (see Sec.~\ref{sec:m78}). If the stellar
UV\footnote{the total number of carbon ionizing photons is
  $6.33~10^{47}~\mathrm{s}^{-1}$} radiation is absorbed by carbon in
the M~78 PDRs, and the kinetic temperature is $T_e = 50~$K, we expect
a flux density of $F_\nu = 2.7$~Jy at 31~GHz for the whole of
M~78. Details are given in \citet[][note there is a typo in their
  Eq.~1, for a missing $D^2$ in the denominator]{cas08}. We use the
recombination coefficients tabulated in \citet{nah97}.

However, \citet{cas08} overlooked the competition of dust and carbon
for the carbon-ionising continuum, which may be estimated by comparing
opacities of dust, $\tau_\mathrm{d}$, and carbon, $\tau_\mathrm{C}$,
to C-ionising radiation. We take a constant photoionisation
cross-section for carbon of $1.6~10^{-17}~$cm$^{2}$ between 11.2 and
13.6~eV, and use the relation $A_\mathrm{V} = N_\mathrm{H} /
2~10^{21}$ \citep[][in CGS units]{boh78}. We find that $\tau_d /
\tau_\mathrm{C} \approx A_\lambda/(A_\mathrm{V} 3.2) \approx 2.0$ for
$\lambda \approx 1000$\AA, for a standard $R_\mathrm{V} = 3.1$, and
using the extinction curves given in \citet{dra03}.  About one third
of all carbon-ionising photons is absorbed by carbon, although this
result is sensitive to the exact value for $R_\mathrm{V}$.

The expected C\,{\sc i} continuum levels at 31~GHz, in a dust-free
nebula, are $F_\nu$ = 2.1, 2.7 and 3.3~Jy, for $T_e$ = 20, 50 and
100~K. But taking into account dust absorption of the carbon-ionising
radiation, these values are reduced by 1/3, to not more than
1~Jy. This, however, is not negligible compared to the observed level
of anomalous emission calculated from the differenced data set, of
$2.79\pm0.81$~Jy.

For a uniform slab nebula, seen face-on as a disk with a radius of
0.1~deg, and  at a distance of 400~pc, the corresponding electron density
at $T_e = 50~$K is 15~cm$^{-3}$, which is $n_\mathrm{H} =
1.5~10^{5}$~cm$^{-3}$ if the carbon in the PDR gas is all singly
ionised. Ionisation balance indicates that the carbon continuum could
reach observable levels, given the observed physical conditions in
M~78 (Sec.~\ref{sec:m78}). However, the contribution of optically thin
carbon continuum at 5~GHz is probably negligible since the 5--31~GHz
index is $\alpha_5^{31} > 0$ (we do not see the nebulae in the PMN
maps). To reach unit opacity at 5~GHz, we require electron densities
of order $\sim 100$~cm$^{-3}$, corresponding to H-nucleus densities of
$\sim 10^6$~cm$^{-3}$ if all of carbon is ionised, with a filling
factor $\sim 10^{-4}$.

Alternatively, the lack of detectable optically thin C\,{\sc i}
continuum at 5~GHz could be used to place constraints on the dust
extinction law in M~78. This requires full-blown PDR models, and is
beyond the scope of this work.


\subsection{Free-free specific intensities} \label{sec:ffintens}

\begin{figure*}
\begin{center}
\includegraphics[width=0.8\textwidth,height=!]{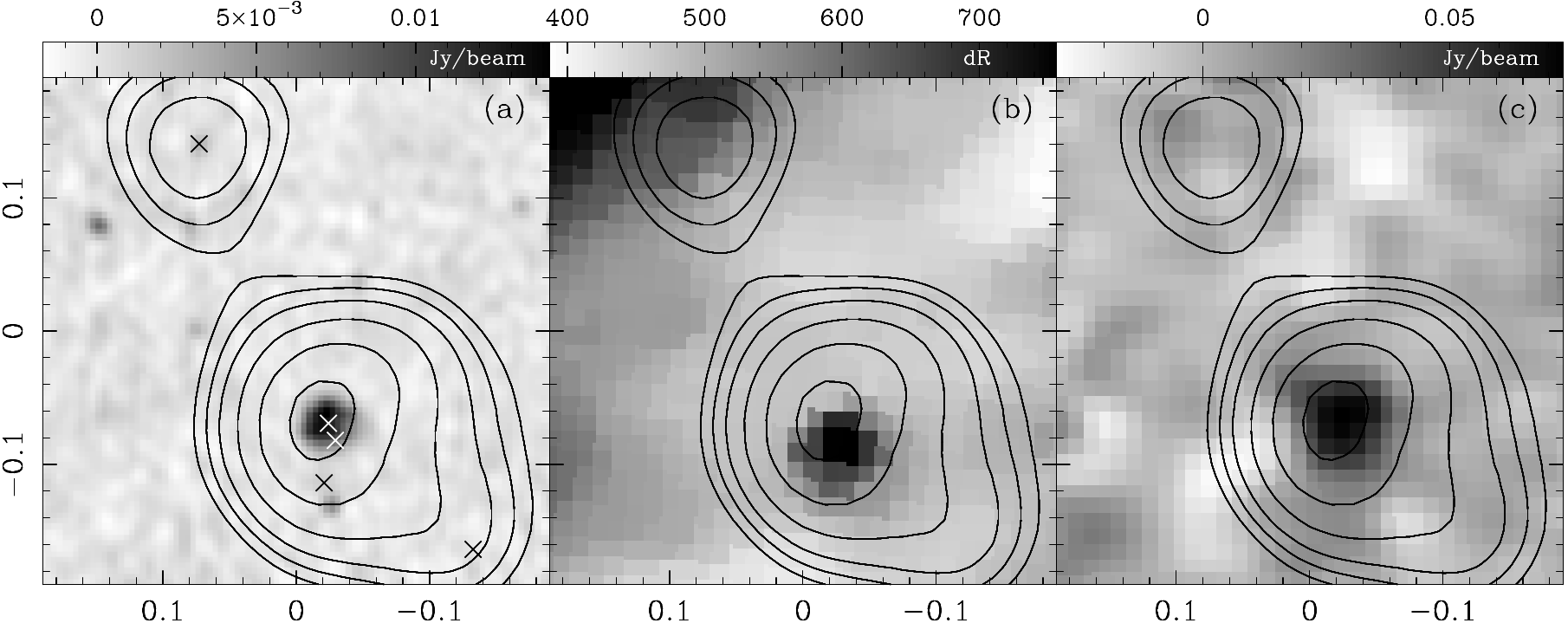}  
\end{center}
\caption{CBI contours over free-free tracer images. The crosses indicate the
  position of the brightest stars, as in Fig.~\ref{figure:corr1}. We find that the
  peaks coincide, but there is no further correlation between CBI and
  any of the templates. (a) CBI contours against NVSS image. (b)
  SHASSA image with CBI contours. In this image we can appreciate the
  extended H$_\alpha$ emission that surrounds M 78. (c) A zoomed image
  from PMN with CBI contours.\label{figure:ff} Coordinates follow
  from Fig.~\ref{figure:large}. }
\end{figure*}

In Fig.~\ref{figure:ff} we compare the CBI contours and the free-free
tracers on small scales.  The 4.85 GHz PMN image traces H\,{\sc i}
free-free. No 4.85~GHz emission is evident at the location of
NGC~2071.  The 4.85 GHz intensity peak, located in NGC~2068, is
$0.073\pm0.011$~Jy~beam$^{-1}$. If we assume an electron temperature
of $T_e = 7000~$K, then the spectral index of optically thin free-free
emission between 5~GHz and 31~GHz is $\alpha = -0.12$
\citep[e.g.][]{wil09,dic03,cas07}, and the peak free-free intensity
should be $\sim 0.055$ Jy beam$^{-1}$ at 31 GHz. H$\alpha$, also a
tracer of H\,{\sc i} free-free, bears similar properties as the PMN
map - no H$\alpha$ is seen in NGC~2071, albeit faint wings from
Barnard's loop.

In each template, i.e. NVSS, SHASSA and PMN, there is compact
free-free emission coincident with the brightest star embedded in
NGC~2068.  It is the compact H\,{\sc ii} region reported by
\citet{mat76}. We see that the peak is coincident with HD38563N for
both CBI and PMN images. 

In order to make a proper comparison between PMN and CBI/CBI2 specific
intensities we have to smooth them to a common resolution. Any
high-pass filtering from PMN will not be significant in this
comparison since, as explained in \citet{con94}, the PMN survey filters-out
scales above 20$\arcmin$, and since the shortest CBI2 baseline is on
20~arcmin scales, CBI2 and PMN have fairly well-matched beams. Being
an interferometer the CBI filters on all scales above its PSF
($\sim$8~arcmin), so we expect the comparison with PMN to give
conservative values in terms of the detection of an excess over
optically thin free-free.  

%
%

The specific intensity units of the PMN data retrievable from
SkyView\footnote{{\tt http://skyview.gsfc.nasa.gov/}} 
are not defined precisely. It is stated in the header of the PMN
images that 200000 beams are about one steradian (so a beam of
7.2~arcmin).  To obtain the exact size of the PMN beam, and to cross
check on the Jy~beam$^{-1}$ units, we fitted a gaussian to 3C273,
which is at a declination similar to that of M~78 (RA
$12^{h}29^{m}06^{s}.7$ DEC $+02\degr 03\arcmin 08\arcsec .6$) so both
images were obtained using similar beams.  Although an elliptical
gaussian fit to 3C273 has a minor axis of 3.8~arcmin FWHM, trials on
neighbouring and fainter radio point sources led to values around
3.4~arcmin.  We assume the original PMN map we are using has a
resolution of 3.7~arcmin. Given the resolution of the CBI maps used
here, of $\sim 5\arcmin .8$, we smoothed the resolution of PMN to that
of CBI by convolving with a gaussian whose FWHM is
$\mathrm{FWHM}_\mathrm{ker}^{2} =
\mathrm{FWHM}_\mathrm{CBI}^{2}-\mathrm{FWHM}_\mathrm{PMN}^{2}$. We
also use this value for the beam units in the PMN image.

%

After smoothing, and scaling Jy~beam$^{-1}$ units to the CBI beam, we
find that the peak intensity in the M~78 PMN image is of $0.015\pm
0.004$~Jy~beam$^{-1}$, while the peak found in the CBI image is of
$0.492\pm 0.031$~Jy~beam$^{-1}$. Using a spectral index $\alpha =
-0.12$ and assuming pure free-free contribution at 4.85~GHz we find an
expected emission of $0.012\pm 0.003$~Jy~beam$^{-1}$ at 31 GHz. The difference
between the actual emission detected at 31~GHz and the one expected from
4.85~GHz free-free is of $0.480\pm0.031$~Jy~beam$^{-1}$. This
corresponds to a spectral index $\alpha = 1.89\pm0.15$ between 4.85~GHz and
31~GHz.

From archival fluxes from PMN and NVSS we find that the peak specific
intensity between 1.4 and 4.85~GHz has a spectral index of
$0.00\pm0.23$. This is an upper limit since NVSS filters out
large scales (NVSS has more flux loss than PMN). We can nonetheless
conclude that the free-free emission seen at 1.4 and 4.85~GHz is
optically thin.


%
%
\subsection{Low Frequency Spectrum of NGC~2071}

Under the hypothesis that the absence of NGC2071 at 5~GHz is due to an
optically thick spectrum, with the observed 31~GHz flux, of
$\sim$0.1~Jy  we expect 16~mJy~beam$^{-1}$ at
5GHz. This is just barely at $2~\sigma$ in PMN, so could be missed.
Also there is hot dust in NGC 2071, as seen in the 25~$\mu$m image,
perhaps due to classical grains heated by a hot star.

For optically thick emission, the expected flux density at 5~GHz from
NGC~2071, 16~mJy~beam$^{-1}$, corresponds to a diameter of
0.05~arcsec.  Such a compact and optically thick source would stand
out in VLA observations at frequencies higher than 5~GHz.
Unfortunately we found no VLA observations that cover the 31~GHz
centroid of NGC~2071.


However, the bulk of the 31~GHz flux density seen towards NGC~2071 is
probably not due to an UCHII region.  The MEM model of the CBI2
visibilities finds structure in NGC~2071, so that its extension is
probably closer to 4\,\arcmin. If opacity is 1 at 31~GHz (so that the
5-31~GHz index is +2), then a distance of 400~pc and an angular
extension of 4\,\arcmin, or a linear depth of 0.5~pc, gives an
electron density of $10^5$~cm$^{-3}$ if $T_e= 10~^4$~K. Such densities
are found only in the most compact UCHII regions, under 0.01~pc in
diameter. In fact no UCHII larger than 0.1~pc are tabulated in
\citet{woo89}, in their Table 16. More recently, \citet{mur10} have
listed UCHII region properties in their Table~3, where an upper limit
of 0.1~pc can be found on the linear sizes of regions with $N_e
\approx 10^5$~cm$^{-3}$. The ionised mass for NGC~2071 would have to
be around 125~M$_\odot$, while the largest UCHII regions reach only
0.5~M$_\odot$. 

\subsection{Radio Stars}

There are several protostars in this region, which contribute to the
total emission at 31~GHz. In most cases the clumps associated with
stellar formation do not have a related radio continuum source
\citep[to a limit of 0.1~mJy at 8.4~GHz][]{gib99}.  Table~2 in
\citet{gib99} gives information on the known 8.4~GHz radio continuum
sources within M~78. We also included a source found in the VLA
archive not included in \citet{gib99}; this sources is roughly
coincident with NGC~2071 IRS~1.

The radio continuum sources will contribute to some extent to the
emission found at 31~GHz. To derive this contribution we use the
formula $S_\nu \propto \nu^\alpha$, with $\alpha$ being the spectral
index. Collimated, ionized stellar winds (sources for radio emission
from young stellar objects) have spectral indeces constrained by $-0.1
< \alpha < 2$, with the highest values corresponding to totally opaque
sources \citep[e.g.][]{rey86}. Using this and considering the highest
possible spectral index we find that the combined contribution from
these radio continuum sources is $237.98\pm11.22$~mJy, far below the
observed emission at 31~GHz which is of $2.73\pm0.28$~Jy. Thus we
neglect any contribution from radio stars to the 31~GHz flux density
in a 45~arcmin aperture.

%

\subsection{Spectral energy distribution fits}

In the previous section we used specific intensity (surface
brightness) to make the comparison between 31 GHz and free-free
templates. Here we use the integrated flux densities to build a SED. For this
we use the values shown in Table~\ref{table:sed} for differenced fluxes,
unless otherwise indicated (See Sec.~\ref{sec:obs_sed}).
We will fit a model of three components of emission: free-free, vibrational
dust and spinning dust.

\begin{figure}
\begin{center}
\includegraphics[width=\columnwidth,height=!]{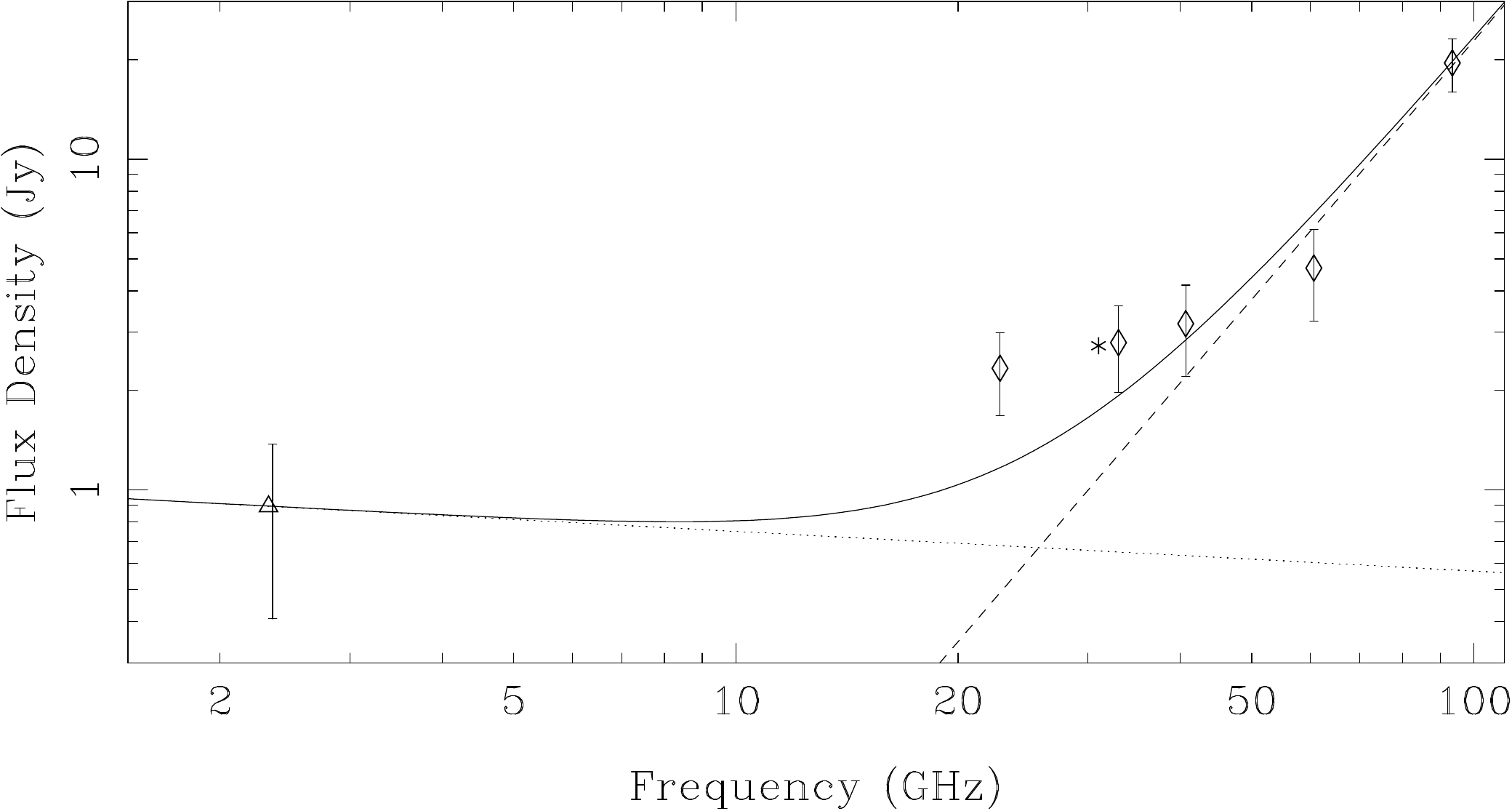}  
\end{center}
\caption{M~78 SED and model components. The spectrum of the radio flux
  in a 45~arcmin aperture is modeled using 2 components made up from
  free-free emission (dotted
  line) and a Rayleigh-Jeans
  tail fitted to the {\em WMAP} data (dashed line). Diamonds
  correspond to flux densities measured by {\em WMAP}, the triangle is
  the flux density obtained by Rhode/HartRAO and the asterisk
  is the flux density from CBI-IRIS 100~$\mu$m correlated emission. The
  solid line is the sum of the three emission components
  considered.\label{figure:sedbeta06}}
\end{figure}

In Fig.\ref{figure:sedbeta06} we see a fit to the data with two
emission mechanisms, not including spinning dust emission. One is free-free
emission for which we used $\alpha = -0.12$, adjusted the datum at
2.326~GHz. We also fitted a modified black-body to the IRIS data points at
100~$\mu$m and 60~$\mu$m. This fit gives us a warm dust temperature of 32~K
and a spectral index $\beta = 1.83$ (we take the spectral index from $S_\nu
\propto \nu^\beta B_\nu(T)$). To check this value we also fitted a modified
black-body to the IRIS images that were CBI-simulated, and found consistent
values of $T = 32$~K and $\beta = 1.82$. We also fitted a Rayleigh-Jeans tail
to the points of {\em WMAP} which accounts for cold dust, assuming that there
is no contribution from spinning dust.  The characteristic values for the
spectral index $\beta$ (defined by $S_\nu \propto \nu^{2+\beta}$) of
ISM dust in the Rayleigh-Jeans regime, ranges from $1 < \beta < 2$
\citep{she09b}. To measure the confidence level of our fit we used a
reduced $\chi^2$ test. For this fit we used a value of $\beta = 0.6$,
lower than characteristic values found in the ISM, but also the
highest value for which the confidence level is $\geq 5$\%. We see
from this fit that a model which uses only Rayleigh-Jeans from cold
dust is unable to explain the emission from M~78 at 95\% confidence.

We used the spinning dust model from \citet{dra98} for dark clouds,
which assumes $n_\mathrm{H} = 10^4$~cm$^{-3}$ and $T_d = 10$~K.  We
fitted this model to the flux densities from differentiated {\em WMAP}
data, except the W band, see Table~\ref{table:sed}. For a given
spinning dust emissivity per H-nucleus (which depends on the physical
environment), the spinning dust intensities are determined by the
column density of hydrogen nuclei, $N_\mathrm{H}$. We fit the {\em
  WMAP} SED by varying $N_\mathrm{H}$, and obtain a value of $(7.99\pm
2.34)~10^{21}$cm$^{-2}$ for $N_\mathrm{H} = f*n_\mathrm{H}*L$, where
$f$ is the filling factor and $L$ is the nebular depth. The
uncertainties on $N_\mathrm{H}$ take into account uncertainties on the
level of free-free and the level of residual Rayleigh-Jeans dust
emission. at 31~GHz.  We see that the cm-wave excess, parametrised by
the spinning dust model, is significant at $\langle N_\mathrm{H}
\rangle = 3.4 \sigma(N_\mathrm{H})$.

The value for $f$ can be constrained using the densities assumed by
the spinning dust model, and those given by \citet{lad97,str75}. We
assume that $n_\mathrm{H}$ ranges from $10^3$ to $10^6$~cm$^{-3}$. We
then consider the nebula as a sphere, with characteristic size of
$27\arcmin$ which is the angular distance from the northeast extreme
NGC~2071 to the southwest end of NGC~2064. When we consider a distance
of 400~pc, we see that $N_\mathrm{H}$ would range from $10^{22}$ to
$10^{25}$~cm$^{-2}$, which constrains $-3.25 < \log(f) < 0.01$. When
using the extinction maps from \citet{sch98} we find H-nucleus column
densities within a smaller range, with $N_\mathrm{H} = (2.28\pm
1.40)~10^{23}$. For this we considered a constant ratio $A_V/E(B-V) =
5.5$ and the relation $N_\mathrm{H} = 10^{22}~A_V/5.3$~cm$^{-2}$
\citep{kim96}. This new range in H-nucleus column densities gives us
the value $\log(f) = -1.46\pm 0.30 $.

%

\begin{figure}
\begin{center}
\includegraphics[width=\columnwidth,height=!]{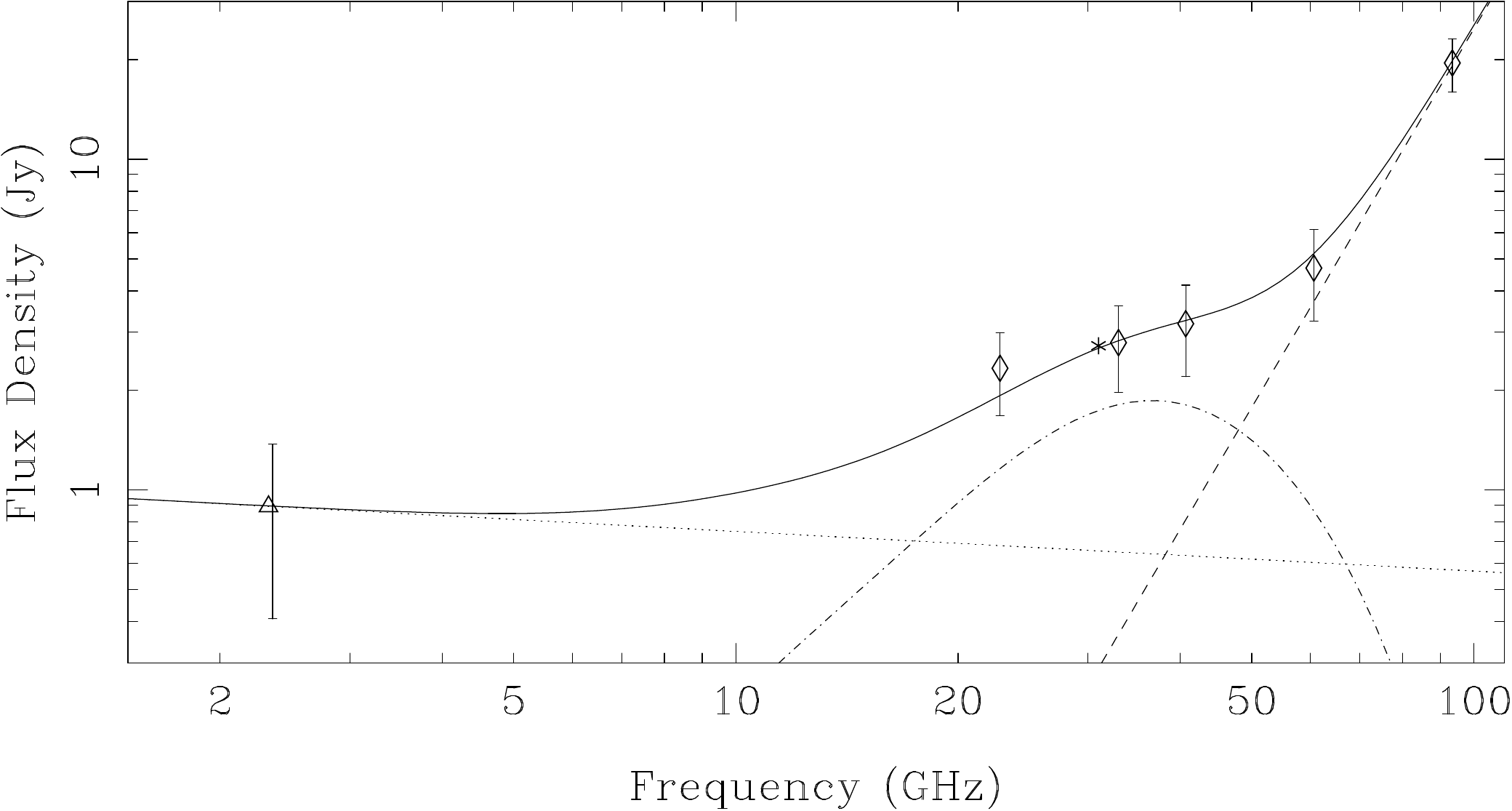}
\end{center}
\caption{M~78 SED and model components. The spectum of the radio flux
  in a 45~arcmin aperture is modeled using 3 components made up from
  the vibrational black body emission of classical dust grains, and
  that from free-free and spinning dust. Diamonds correspond to flux
  densities measured by {\em WMAP}, the triangle is the flux density
  obtained by Rhodes/HartRAO and the asterisk is the flux density from
  CBI-IRIS 100~$\mu$m correlated emission. The dotted line is the
  free-free spectrum derived from 2.326~GHz, the dashed line
  corresponds to the sub-mm dust tail, traced at 94~GHz and the
  dash-dotted line represents  a spinning dust model. The solid line
  is the sum of the three emission components
  considered.\label{figure:sed}}
\end{figure}

Fig.~\ref{figure:sed} shows the spectral energy
distribution calculated from {\em WMAP}, Rhodes/HartRAO and CBI, this last one
extracted from IRIS 100~$\mu$m using the coefficient obtained from the
visibility plane correlation (see Sec.~\ref{sec:obs_sed}). The free-free
emission was derived from the datum at 2.326 GHz using a spectral index
$\alpha = -0.12$. We assume that the excess in
{\em WMAP} at 94 GHz from the modified black-body fitted using IRIS at 100
and 60~$\mu$m traces the sub-mm cool dust grey-body. We fixed to this
datum point the Rayleigh-Jeans tail of vibrational dust using a characteristic
$\beta = 1.8$ which lies within the range established by \citet{she09b}.

We assume that spinning dust is traced by dust-correlated radio
emission, using the IRIS 100~$\mu$m template, so that the 31 GHz
emission is given by $F_{31 \mathrm{GHz}} = a F_{100 \mu\mathrm{m}}$,
where $a = (0.092\pm0.001)~10^{-3}$ is the slope of the correlation in
the visibility plane. If all visibilities were weighted equally, this
would be equivalent to correcting the CBI flux density for flux loss.

If we again use 2.326~GHz to trace free-free emission, but this time
we differentiate to remove the diffuse background, we find that the
free-free level is  $0.65\pm0.35$~Jy at 31~GHz. Compared with the
value from dust-correlated emission at 31~GHz, which is of
$2.73\pm0.28$~Jy, we have an excess of $2.08\pm0.45$~Jy over the expected
level of optically thin H\,{\sc i} free-free.  The emission at 31~GHz is not
dominated by free-free, confirming the results from Sec.~4.5.

Part of the Rayleigh-Jeans tail of sub-mm dust emission could reach
31~GHz. Fixing the level of the Rayleigh-Jeans flux density to the
point at 93.5~GHz, with $\beta = 1.6\pm0.3$ (which extends
conservatively to shallow values), we expect $0.40\pm0.19$~Jy at
31~GHz. We see that the sub-mm dust cannot account for the observed
100~$\mu$m-correlated flux density at 31~GHz of $2.73\pm0.28$~Jy
(close to the 33~GHz flux density from {\em WMAP}~33~GHz, of
$2.79\pm0.81$). When we consider the addition of vibrational dust
emission and free-free emission, as previously derived, we find a
31~GHz excess of $F_{31~\mathrm{GHz}}=1.68\pm0.49$~Jy, significant at
3.4~$\sigma$.  Comparing this value with the expected $2.73\pm0.28$~Jy
we can say that at least $62\pm19$\%, of the observed 31~GHz emission
cannot be explained by H\,{\sc i} free-free and/or vibrational dust
emission. 

The addition of a spinning dust component improves significantly the
confidence level (as derived from a reduced $\chi^2$ test) for the SED fit,
when compared with models that only consider the Rayleigh-Jeans tail from sub-mm
dust to account for the spectra at 20-60~GHz. The only free parameter
for a spinning dust component is $f*N(\mathrm{H})$, whose value of $(7.99\pm
2.34)~10^{21}$cm$^{-2}$ differs significantly from zero. From this we can say
that spinning dust is a significant component in our SED for M~78 at angular
scales of $\sim 8\arcmin$.

\subsection{Comparison with other radio - IR proportionality
  coefficients} 

For comparison with previous work on the radio-IR correlations in
other objects, we converted our correlation slopes to
$T_{31~\mathrm{GHz}}/I_{100~\mu\mathrm{m}}$, i.e.  $\mu$K~(MJy~sr$^{-1}$)$^{-1}$
instead of dimensionless units\footnote{the radio-IR correlation
  slopes in units of $\mu$K~(MJy~sr$^{-1}$)$^{-1}$ are sometimes referred
  to as `emissivities'}. These values are summarized in
Table~\ref{table:slopes}. We found that the highest slopes correspond to the
fainter nebulae, NGC~2064 and NGC~2067, with values of $7.69\pm0.96$ and
$7.21\pm1.72$~$\mu$K~(MJy~sr$^{-1}$)$^{-1}$, while for NGC~2071 and NGC~2068
we found $4.51\pm0.14$ and $4.29\pm0.25$~$\mu$K~(MJy~sr$^{-1}$)$^{-1}$. We see
that all these values are within the range constrained by
$3.3\pm1.7$~$\mu$K~(MJy~sr$^{-1}$)$^{-1}$, derived by \citet{dic07}
from 6 H\,{\sc ii} regions, and the value of
$11.2\pm5.0$~$\mu$K~(MJy~sr$^{-1}$)$^{-1}$ for cool dust obtained from
15 regions by \citet{dav06}. Regions with higher
temperatures tend to have lower slopes.  The radio-correlated dust
emission stems from hotter dust in H\,{\sc ii} regions than in
M~78. Also we see that the correlation slopes in  NGC~2071 and
NGC~2068 are lower than in NGC~2067 and NGC~2064, as expected since
the exciting stars in NGC~2071 and NGC~2068 are hotter.

\begin{table}
\caption{\label{table:slopes} Radio-IR correlation slopes  for individual
  clouds in M~78.}
\begin{tabular}{ccrrrrrr}
Source &  $\mu$K~(MJy~sr$^{-1}$)$^{-1}$ \\ \hline
 NGC~2068 & $4.29\pm 0.25$ \\
 NGC~2071 & $4.51\pm 0.14$ \\
 NGC~2067 & $7.21\pm 1.72$ \\
 NGC~2064 & $7.69\pm 0.96$ \\
 6 H\,{\sc ii} regions$^{a}$ & $3.3\pm1.7$ \\
 15 Cool dust regions$^{b}$ & $11.2\pm 5.0$ \\ \hline
\end{tabular}
\begin{flushleft}
$^{a}${\citet{dic07}.} $^{b}${\citet{dav06}.}
\end{flushleft}
\end{table}

\section{Conclusions} \label{sec:conc}


We have found evidence of dust-correlated emission towards M~78 at a
frequency of 31~GHz. Its spatial distribution is coincident with that
from dust, especially as traced at 12 and 100~$\mu$m, for which we
found similar linear correlation coefficient in the visibility
plane. The morphology at 31~GHz is qualitatively inconsistent with
pure optically thin free-free emission, as can be traced by PMN, NVSS
and SHASSA, where we only find significant emission around HD38563N.

The radio-IR correlations worsen at finer angular resolutions. The
CBI2 and IRAC~8~$\mu$m correlation tests worsen relative to those
found from CBI and IRIS. The correlation tests further degrade when
dividing the IR images by the UV-field. Thus the radio/IR differences
cannot be explained as being due to modulation by the UV-field.

%

The specific intensity levels seen by CBI at 31~GHz are higher than those
expected for free-free emission derived from PMN data (smoothed to the CBI
resolution) by $0.480\pm0.031$~Jy~beam$^{-1}$. However, the flux
densities measured in a 45~arcmin circular aperture are dominated by diffuse
free-free emission, as suggested by flux measurements with {\em WMAP}
and Rhodes/HartRAO. When we differentiate against a reference field
within the Orion B complex, we find an excess flux density at 31~GHz
of $2.08\pm0.52$~Jy over the expected free-free in a 45~arcmin
circular aperture. We need high resolution data to subtract more
accurately the background free-free emission and confirm the values
found in this work.

%

When fitting a Rayleigh-Jeans law to {\em WMAP} differentiated fluxes
we derive a spectral index $\beta = 0.6$. This value is lower than
those typically found in the ISM, where $1 < \beta < 2$, but even
allowing for this value of $\beta$ the confidence level of this model
spectrum, given the observed SED, is only of 5~\%.  We can neglect the
contribution of a Rayleigh-Jeans tail from sub-mm dust at
cm-wavelengths.

The addition of a spining dust component to the SED fit results in a
significant improvement in confidence level. The only free parameter
in our fit, the column of H-nuclei which scales the model spinning
dust emissivities per H-nucleus, differs significantly from zero,
thus confirming that the higher confidence level achieved is due to a
non negligible component. This is also supported by the fact that the
combined contributions of Rayleigh-Jeans and free-free at 31~GHz yield a
total flux that is $1.68\pm0.49$~Jy below the emission detected by CBI. A
spinning-dust-like component is needed in order to explain the observed
emission at cm-wavelengths in M~78. 

%

\section*{Acknowledgments}

We thank Evelyne Roueff for interesting discussions on PDRs.
S.C. acknowledges support from a Marie Curie International Incoming
Fellowship (REA-236176), from FONDECYT grant 1100221, and from the
  Chilean Center for Astrophysics FONDAP 15010003. CD acknowledges an
  STFC Advanced Fellowship, and a ERC grant under the FP7.  This work
  has been supported in part by the Strategic Alliance for the
  Implementation of New Technologies (SAINT - see {\tt
    www.astro.caltech.edu/chajnantor/saint/index.html}). This work has
  been carried out within the framework of a NASA/ADP ROSES-2009
  grant, n. 09-ADP09-0059. The CBI observations were made possible
  thanks to the engineering team: C. Achermann, R. Bustos, C. Jara,
  N. Oyarce, R. Reeves, M. Shepherd and C. Verdugo. We acknowledge the
  use of the Legacy Archive for Microwave Background Data Analysis
  (LAMBDA). Support for LAMBDA is provided by the NASA Office of Space
  Science. We used data from the Southern H-Alpha Sky Survey Atlas
  (SHASSA), which is supported by the National Science Foundation.

\label{lastpage}

\end{document}